\documentclass[twocolumn]{aa}

\usepackage{graphicx}
\usepackage{epstopdf}
\usepackage{deluxetable}
\usepackage{times,epsfig} 

\usepackage{amssymb}
\usepackage{mathrsfs} 
\usepackage{dirtytalk}
\usepackage{acronym}

\usepackage{natbib}
\bibpunct{(}{)}{;}{a}{}{,} 
\usepackage{caption}

\usepackage{amsmath}
\usepackage[english]{babel}
\usepackage{longtable}
\usepackage{afterpage}
\usepackage{url}
\usepackage{hyperref}
\usepackage{xcolor}
\definecolor{dark-red}{rgb}{0.4,0.15,0.15}
\definecolor{dark-blue}{rgb}{0.15,0.15,0.4}
\definecolor{medium-blue}{rgb}{0,0,0.5}
\hypersetup{
    colorlinks, linkcolor={dark-red},
    citecolor={dark-blue}, urlcolor={medium-blue}
}

\acrodef{ztf}[ZTF]{Zwicky Transient Facility}
\acrodef{lc}[LC]{light curve}

\newcommand{\beqa}{\begin{eqnarray}} 
\newcommand{\eeqa}{\end{eqnarray}}

\newcommand{\bsub}{\begin{subequations}}
\newcommand{\esub}{\end{subequations}}
\newcommand{\beal}{\begin{align}}
\newcommand{\ealn}{\end{align}}

\newcommand{\Msun}{{\ensuremath{\mathrm{M}_{\odot}}}}

\newcommand{\swift}{\textit{Swift}}

\begin{document}
\title{Three Core-Collapse Supernovae with Nebular Hydrogen Emission.}

\author{J. Sollerman\inst{1} % \author[0000-0003-1546-6615]{J.~Sollerman}
\and{S. Yang}\inst{1}
\and{S. Schulze}\inst{1} 
\and{N. L. Strotjohann}\inst{2}
\and{A. Jerkstrand}\inst{1}
\and{S.~D.~Van Dyk}\inst{3} 
\and{E.~C. Kool}\inst{1}
\and{C. Barbarino}\inst{1,4}
\and{T. G. Brink}\inst{5}
\and{R. Bruch}\inst{2}
\and{K. De}\inst{6}
\and{A. V. Filippenko}\inst{5,7}
\and{C. Fremling}\inst{6}
\and{K. C. Patra}\inst{5}
\and{D. Perley}\inst{8}
\and{L. Yan}\inst{6}
\and{Y. Yang}\inst{5}
\and{I. Andreoni}\inst{6} 
\and{R. Campbell}\inst{9}
\and{M. Coughlin}\inst{10}
\and{M. Kasliwal}\inst{6} 
\and{Y.-L. Kim}\inst{11}
\and{M. Rigault}\inst{11}
\and{K. Shin}\inst{7}
\and{A. Tzanidakis}\inst{6}
\and{M. C. B. Ashley}\inst{12}
\and{A. M. Moore}\inst{13}
\and{T. Travouillon}\inst{13}
}
\institute{Department of Astronomy, The Oskar Klein Center, Stockholm University, AlbaNova, 106 91 Stockholm, Sweden
\and{Department of Particle Physics and Astrophysics, Weizmann Institute of Science, 234 Herzl St, 76100 Rehovot, Israel}
\and{Caltech/IPAC, Mailcode 100-22, 1200 E.~California Blvd., Pasadena, CA 91125, USA}
\and{The Raymond and Beverly Sackler School of Physics and Astronomy, Tel Aviv University, Tel Aviv, 69978 Israel}
\and{Department of Astronomy, University of California, Berkeley, CA 94720-3411, USA}
\and{Cahill Center for Astrophysics, California Institute of Technology, 1200 E. California Blvd. Pasadena, CA 91125, USA}
\and{Miller Institute for Basic Research in Science, University of California, Berkeley, CA 94720, USA}
\and{Astrophysics Research Institute, Liverpool John Moores University, Liverpool Science Park, 146 Brownlow Hill, Liverpool L35RF, UK }
\and{W. M. Keck Observatory, Kamuela, HI 96743, USA}
\and{School of Physics and Astronomy, University of Minnesota, Minneapolis, MN 55455, USA}
\and{University Claude Bernard Lyon 1, CNRS, IP2I Lyon / IN2P3, IMR 5822, F-69622, Villeurbanne, France}
\and{School of Physics, University of New South Wales, NSW 2052, Australia}
\and{Research School of Astronomy and Astrophysics, Australian National University, Canberra, ACT 2611, Australia}
}

%\date{Received; Accepted}

\abstract
{  {
We present the discovery and extensive follow-up observations of SN\,2020jfo,
a Type IIP supernova (SN) in the nearby (14.5 Mpc) galaxy M61. 
Optical light curves (LCs) and spectra from the Zwicky Transient Facility (ZTF),
complemented with data from {\it Swift}/UVOT and
near-infrared photometry is presented. These are used to model the
350-day duration bolometric light curve, which 
exhibits a relatively short ($\sim 65$ days) plateau.
This implies a moderate ejecta mass ($\sim 5$ \Msun) at the time of explosion,
whereas the deduced amount of ejected radioactive nickel is $\sim 0.025$ \Msun.

An extensive series of spectroscopy is presented, including spectropolarimetric observations. The nebular spectra are dominated by H$\alpha$ but also reveal emission lines from oxygen and calcium. Comparisons to synthetic nebular spectra indicate an initial progenitor mass of $\sim 12$ \Msun. We also note the presence of stable nickel in the nebular spectrum, and SN\,2020jfo joins a small group of SNe that have inferred super-solar Ni/Fe ratios.

Several years of pre-discovery data are examined, but no signs of pre-cursor activity is found. Pre-explosion {\it Hubble Space Telescope} imaging reveals a probable progenitor star, 
detected
only in the reddest band ($M_{\rm F814W} \approx -5.8$) and is fainter than expected for stars in the 10 -- 15 \Msun range. There is thus some tension between the LC analysis, the nebular spectral modeling and the pre-explosion imaging. 

To compare and contrast, we present two additional core-collapse SNe monitored by the ZTF, which also have nebular H$\alpha$-dominated spectra. This illustrates how the absence or presence of interaction with circumstellar material (CSM) affect both the LCs and in particular the nebular spectra.
Type~II SN\,2020amv has a LC powered by CSM interaction, in particular after $\sim40$ days when the LC is bumpy and slowly evolving. The late-time spectra show strong H$\alpha$ emission with a structure suggesting emission from a thin, dense shell. The evolution of the complex three-horn line profile is reminiscent of that observed for SN\,1998S. Finally, SN\,2020jfv has a poorly constrained early-time LC, but is of interest because of the transition from a hydrogen-poor Type IIb to a Type IIn, where the nebular spectrum after the light-curve rebrightening is dominated by H$\alpha$, although with an intermediate line width.
}}


\authorrunning{Sollerman et al.}
\titlerunning{Three different ZTF CC SNe with nebular H$\alpha$.}
\maketitle

\section{Introduction}
\label{sec:intro}

Core-collapse (CC) supernovae (SNe) are explosions of massive stars ($\gtrsim8~M_\odot$) ending their stellar lives. 
The variety of CC SNe is determined primarily by the progenitor mass at the time of CC, but also by the mass-loss history leading up to the explosion. 
The most common category is the hydrogen-rich class; Type II SNe.
Hydrogen-poor CC SNe also originate from massive progenitor stars, but stars that have lost most --- or even all --- of their H envelopes prior to explosion; 
Type IIb SNe (some H left), SNe Ib (no H, some He), and SNe Ic (neither H nor He); see 
\cite{Filippenko1997} and \cite{galyam2017hsn..book..195G}
for reviews.
There are few observational constraints on mass loss for massive stars, and the processes involved are poorly understood, but it is well established that 
the interaction between the ejecta and the circumstellar material (CSM) can make a significant contribution to the total SN luminosity \citep[e.g.,][]{ChevalierFransson2017}.

In this paper, the primary aim is to present the discovery and follow-up observations of a particularly nearby supernova, the relatively normal Type II SN 2020jfo in the grand-design spiral galaxy M61, only 14.5 Mpc away. We present the discovery by the Zwicky Transient Facility (ZTF) of this transient and the optical light curves (LCs), as well as a spectral sequence covering the first 350 days of its evolution. 
For comparison, we include two additional ZTF SNe that (like SN 2020jfo) are dominated by strong H$\alpha$ emission in their nebular spectra.
The Type II SN 2020amv has a LC dominated by CSM interaction, which also reveals itself in the complex line profiles seen in the nebular spectrum. SN 2020jfv was originally classified as a Type IIb SN (i.e. a relatively hydrogen-poor transient). However, the LC at later times starts rebrightening and the nebular spectrum is also dominated by H$\alpha$, with a distinct emission-line profile.  These three objects from the ZTF survey are thus used to exemplify the appearance of the nebular Balmer lines, and how these are connected to different LC shapes and powering mechanisms.

The paper is organised as follows. 
In Sect.~\ref{sec:obs} we present the observations, including optical photometry and spectroscopy as well as 
near-infrared (NIR) photometry and
near-ultraviolet (UV) space-based data. 
Section~\ref{sec:discussion} presents the similarities and differences between the objects, and Sect.~\ref{sec:conclusions} summarises our conclusions 
and discusses our observations in context with other SNe.

%==========================================
\section{Observations and Reductions}
\label{sec:obs}
%==========================================

\subsection{Detection and classification}
\label{sec:detection}

\subsubsection{SN 2020jfo}

SN\,2020jfo (also known as ZTF20aaynrrh)
was discovered on 
2020 May 6 (UT dates are used throughout this paper; first detection on $\mathrm{JD}=2458975.70$), with the Palomar Schmidt 48-inch (P48) Samuel Oschin telescope as part of the ZTF survey \citep{Bellm2019PASP..131a8002B,GrahamM2019}. It was reported to the Transient Name Server 
(TNS\footnote{\href{https://wis-tns.weizmann.ac.il}{https://wis-tns.weizmann.ac.il}})
on the same day
\citep{2020TNSTR1248....1N}, 
less than 2~hours after first detection. 
The first detection is in the $r$ band, 
with a host-subtracted magnitude of $16.01\pm0.04$, at the J2000.0 coordinates $\alpha=12^{h}21^{m}50.48^{s}$, $\delta=+04\degr28\arcmin54.1\arcsec$. 
The first report also mentions that the last upper limit ($g > 19.7$ mag) from ZTF was on May 2, 4 days before discovery.
Following \cite{Bruch_2021}, we perform a power-law fit to the
early-phase $g$ and $r$ light curves, and obtain an estimated
explosion date of
$\mathrm{JD_{explosion}^{SN2020jfo}}=2458975.20$. 
A conservative uncertainty is $^{+0.5}_{-3.5}$ days as provided by the first detection and the last nondetection.
We use this as the explosion date throughout the paper, and measure the phases in rest-frame days with respect to it.
This transient was subsequently also reported to the TNS by several other surveys (for example, by ATLAS and PS2 in May and by {\it Gaia} and MASTER in June), but was also bright enough to be followed by many amateur astronomers around the 
globe\footnote{\href{https://www.rochesterastronomy.org/sn2020/sn2020jfo.html}{https://www.rochesterastronomy.org/sn2020/sn2020jfo.html}}.

\begin{figure*}
\centering
     \includegraphics[width=0.8\textwidth]{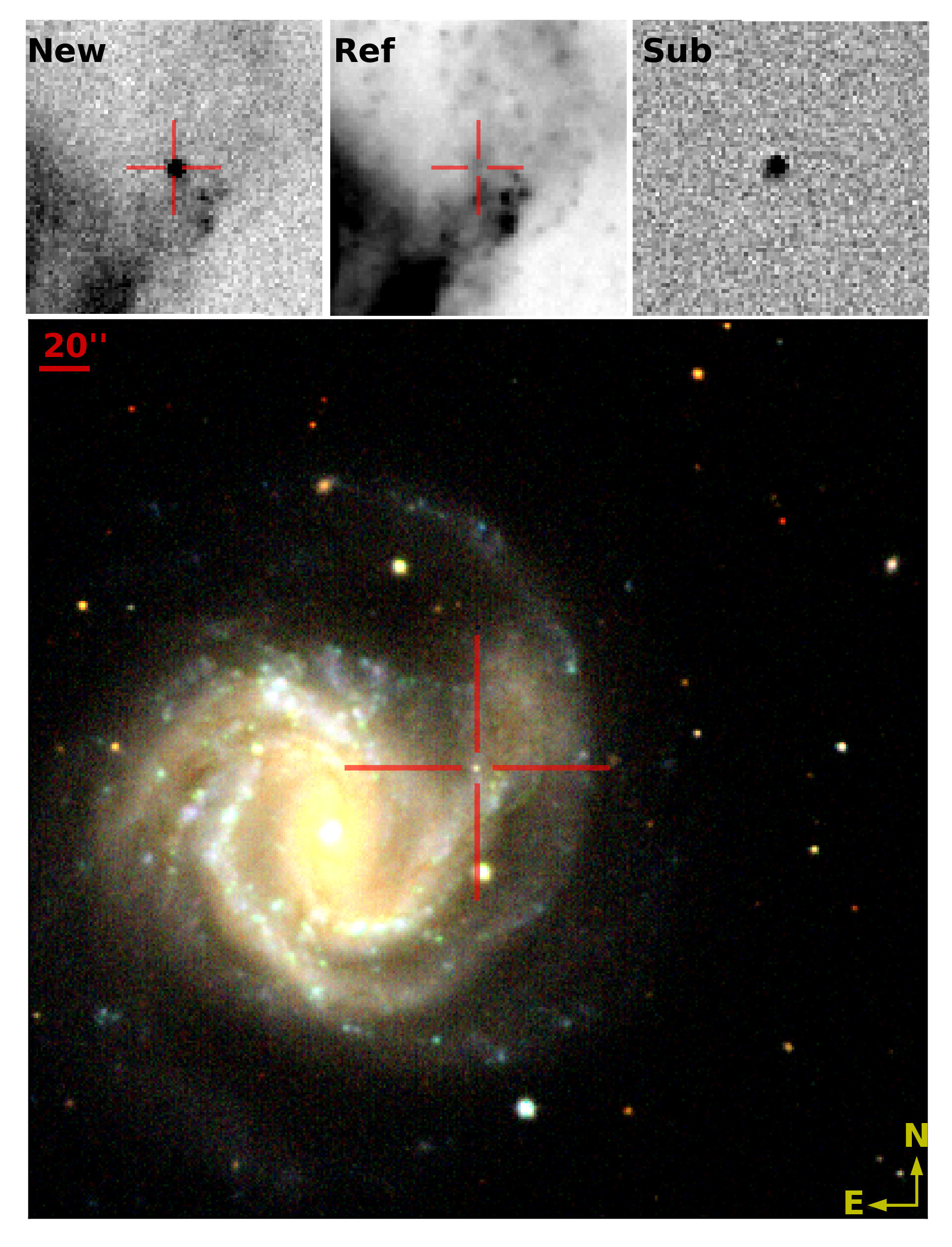}
    \caption{SN 2020jfo in M61.
    The $r$-band image subtraction is shown in the top panels, with the SN image observed 
    on 2020 May 6 on the left; this is the \ac{ztf} discovery frame. The SN position is marked. The middle top panel is the reference image (the template), and the right panel shows the SN standing out in the image subtraction.
    The bottom panel displays a $gri$-colour composite image of the host galaxy and its environment.
    It was composed of \ac{ztf} $g$, $r$, and $i$ images of the field observed on 2020 December 21, $\sim 7$ months after the first \ac{ztf} detection. {  The SN is still visible and marked by the red cross.} }
    \label{fig:image}
\end{figure*}

SN 2020jfo is positioned in the spiral galaxy
M61 (NGC 4303), which has a redshift of
$z = 0.00522$. 
It lies in the Virgo cluster and has hosted 7 known SNe before SN 2020jfo, the most recent being the Type Iax SN 2014dt \cite[e.g.,][]{2018PASJ...70..111K}.
As is often the case, the distance of this nearby host is relatively uncertain. We follow \citet[][]{2018PASJ...70..111K} and adopt 
a distance modulus of $30.81\pm0.20$ mag, which is 14.5 Mpc. {  The implications for the uncertainty in the distance is discussed in Sect.~\ref{sec:uncertainties}}.
The position of SN 2020jfo in M61 is shown in Fig.~\ref{fig:image}.

The ZTF on-duty astronomer (J.S.) who first noticed the SN immediately triggered the robotic 
Palomar 60-inch telescope (P60; \citealp{2006PASP..118.1396C}) equipped with the 
Spectral Energy Distribution Machine (SEDM; \citealp{2018PASP..130c5003B}). 
Unfortunately, the observations could not be scheduled the same night, so we instead triggered observations from La Palma. 
We thus classified SN 2020jfo \citep{2020TNSCR1259....1P} as a Type II SN
based on a spectrum obtained on 2020 May 6 with the
Liverpool telescope (LT) equipped with the SPectrograph for the Rapid Acquisition of Transients (SPRAT),
as well as on a spectrum from the 
Nordic Optical Telescope (NOT) using the Alhambra Faint Object Spectrograph (ALFOSC).
These spectra were obtained 
17.27 and 17.55~hr after the first ZTF detection.
The first SEDM spectrum came in a few hours thereafter, 
and corroborated the classification
\citep{2020TNSAN.100....1P}.

\subsubsection{SN 2020amv}
Our first ZTF photometry of SN 2020amv (ZTF20aahbamv)
was obtained on 2020 January 23 ($\mathrm{JD}=2458871.72$) with the P48.
It was saved by the on-duty astronomer (J.S.)
to the GROWTH Marshal \citep{Kasliwal2019}.
The first detection was in the $g$ band, with a host-subtracted magnitude of $18.68\pm0.08$, at $\alpha=08^{h}49^{m}40.68^{s}$, $\delta=+30\degr11\arcmin14.5\arcsec$ (J2000.0).
The source was reported to TNS on the same day  
\citep{2020TNSTR.239....1N}, 
with a note saying that the latest nondetection from ZTF was 4 days prior to discovery. 
This was also a ZTF discovery, but ATLAS reported the same object just a few hours later. 
We include the forced photometry LCs from ATLAS \citep{2018PASP..130f4505T,2020PASP..132h5002S}
when available, for completeness.  
With power-law fits to the early $g$ and $r$ data, we set the explosion date as 
$\mathrm{JD_{explosion}^{SN2020amv}}=2458871.22\pm0.29$. 

The transient was classified as a Type II SN by ePESSTO+ \citep{smarttpessto}
using a spectrum from February 2 \citep{2020TNSCR.369....1I}, and the SEDM spectrum from the Bright Transient Survey
{  \citep{FremlingBTS2020ApJ...895...32F}}
was also made public on TNS \citep{2020TNSCR1491....1D}.
SN 2020amv was observationally well covered at early phases 
since it showed narrow features that could indicate CSM interaction (``flash features"; see, e.g., \citealt[][]{GalYam2014,Bruch_2021}). The host galaxy has $z = 0.0452$, and using a flat $\Lambda$CDM cosmology with $\Omega_{\rm m}=0.3$ and H$_{0} = 70$~km~s$^{-1}$~Mpc$^{-1}$ this corresponds to 200 Mpc (distance modulus 36.5 mag). 

\subsubsection{SN 2020jfv}
SN\,2020jfv (ZTF20abgbuly) was first reported to TNS by ATLAS
\citep{2020TNSTR1243....1T}, with a detection on 2020 May 5. 
{\it Gaia} actually detected the transient earlier (on April 30), and claimed a nondetection the previous night.
With ZTF the first observations were obtained later (June 18), when the object was already clearly declining.
For SN 2020jfv we adopt 2020 April 30 
($\mathrm{JD_{explosion}^{SN2020jfv}}=2458969.51$)  
as both the discovery date and the explosion date, 
but note that there is quite some uncertainty in the actual date of explosion.
The object is positioned at
$\alpha=23^{h}06^{m}35.75^{s}$, $\delta=+00\degr36\arcmin43.7\arcsec$ (J2000.0),
not far from the centre of the face-on spiral galaxy WISEA J230635.97+003641.9. 
This galaxy had no previously reported redshift, and our estimate of $z=0.017$ comes from the measured
host-galaxy narrow emission lines in our late-time nebular spectra.
The distance is thus estimated to be 
73.8
Mpc adopting the same cosmology as provided above.
The SN was classified as Type IIb based on an SEDM spectrum obtained on June 20
 \citep{2020TNSCR1948....1D}.
It continued to fade in a linear fashion for the next 100 days, but it thereafter began rebrightening, especially in the $r$ band. Late-time spectra demonstrate that this is due to CSM interaction driving a conspicuous nebular H$\alpha$ line. The transformation from a stripped-envelope SN to a CSM interacting Type IIn is reminiscent of the evolution discussed by {  e.g., \citet{sollerman2020,milisavljevic2015,mauerhan2018,prentice2020,chandra2020,Tartaglia2021}}.
Some of the data for these three SNe and their host galaxies is summarised in Table~\ref{table:SNproperties}.

\subsection{Optical photometry}
\label{sec:optical}

Following the discoveries as outlined above, ZTF obtained regular follow-up photometry in the $g$, $r$, and sometimes $i$ bands with the ZTF camera 
\citep{dekany2020} on the P48.
Additional images were obtained with the LT and with the P60.
Some late-time photometry was also obtained with ALFOSC on the NOT.
P48 LCs come from the ZTF pipeline \citep{2019PASP..131a8003M}, where we have also applied forced photometry \citep[see, e.g.,][]{yuhan2019}.
Photometry from the P60 and LT was produced with the image-subtraction pipeline described by \cite{Fremling2016}, with template images from the Sloan Digital Sky Survey (SDSS; \citealp{2014ApJS..211...17A}). This pipeline produces point-spread-function (PSF) magnitudes, calibrated against SDSS stars in the field. 
The NOT photometry was done using 
template subtractions performed with {\tt hotpants}\footnote{\url{ http://www.astro.washington.edu/users/becker/v2.0/hotpants.html}}, using archival SDSS images as templates.
The magnitudes of the transient were measured using SNOoPY\footnote{SNOoPy is a package for SN photometry using PSF fitting and template subtraction developed by E. Cappellaro. A package description can be found at \url{http://sngroup.oapd.inaf.it/snoopy.html}.} and calibrated against SDSS stars in the field.
All magnitudes are reported in the AB system 
\citep{okeandgunn1983}.
For completeness and comparison, we have also included data from the forced photometry service from ATLAS. 
Those data points are included in the figures, but are generally not part of the analysis and measurements.

\begin{figure*}
\centering
     \includegraphics[width=\textwidth]{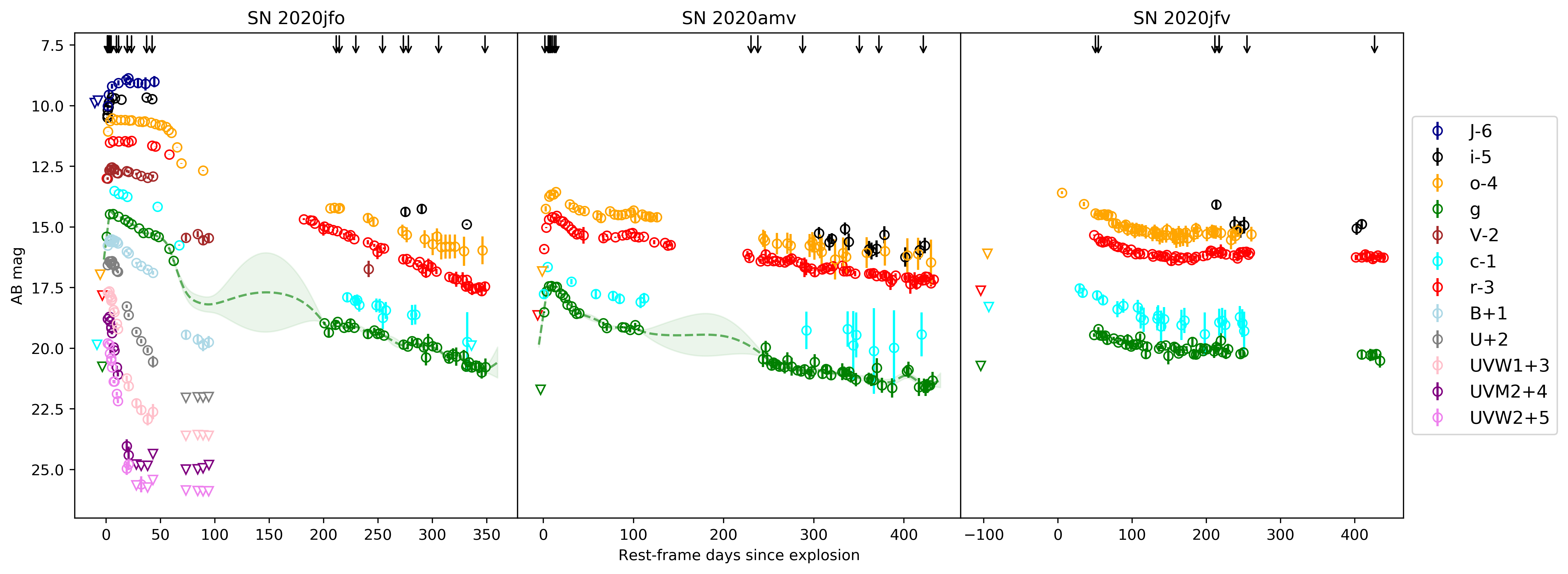}
    \caption{Light curves of the three supernovae. 
    Left: SN 2020jfo in multiple bands with symbols and offsets as provided in the legend to the very right. In this case it includes six passbands from \swift\ and three from ZTF and LT, as well as near-infrared $J$-band data. 
    These are observed (AB) magnitudes plotted versus rest-frame time in days since explosion. 
    For the $J$-band data, the Vega/AB magnitude conversion follows \cite{Blanton2007}.
    ATLAS forced photometry in the $c$ and $o$ bands is also included, and the ZTF and ATLAS data are in 3 day bins. The arrows on top indicate the epochs of spectroscopy, and the lines with error regions are Gaussian Process estimates of the interpolated LCs (for $g$ band). The middle and right panels show the LCs for SN 2020amv and SN 2020jfv, respectively.
    }
    \label{fig:lcs}
\end{figure*}

In our analysis we have corrected all photometry for Galactic extinction, using the Milky Way (MW) colour excess 
$E(B-V)_{\mathrm{MW}}$ toward the position of the SNe, as provided in Table~\ref{table:SNproperties}.
All reddening corrections are applied using the \cite{1989ApJ...345..245C} extinction law with $R_V=3.1$. 
No further host-galaxy extinction has been applied. {  We discuss the effects of this assumption in Sect.~\ref{sec:uncertainties}}.
All the photometry is given in 
Tables~\ref{tab:2020jfo_phot}, \ref{tab:2020amv_phot} and \ref{tab:2020jfv_phot}.
The three light curves are shown in Fig.~\ref{fig:lcs}.

\subsection{Near-infrared photometry}
\label{sec:NIRphot}

SN 2020jfo was observed in the NIR $J$ band as a part of regular survey operations of the Palomar Gattini-IR (PGIR) survey. PGIR is a wide-field NIR survey at Palomar Observatory, observing the entire visible sky at a median cadence of $\sim 2$\,days, and to a median 5$\sigma$ depth of $J \approx 15.7$\,AB mag \citep{Moore2019, De2020a}. The transient was clearly detected in the PGIR data for $\sim 30$\,days after discovery.
We derived $J$-band photometry (calibrated to the 2MASS catalogue in the Vega system) of SN 2020jfo by performing forced PSF photometry at the location of the transient on the PGIR difference images, using the method described by \citet{De2020b}. The photometry is given in Table~\ref{tab:pgir} 
and included in Fig.~\ref{fig:lcs}.

\subsection{Swift observations}
\subsubsection{UVOT photometry\label{sec:uvot}}

For SN 2020jfo, we also have
observations in the 
ultraviolet (UV) and optical from the UV Optical Telescope onboard the Neil Gehrels
\swift\ observatory (UVOT; \citealp{2004ApJ...611.1005G}; \citealp{2005SSRv..120...95R}). 
As shown in Table~\ref{tab:swift}, 25 epochs were obtained over the first 300 days.
Our first \swift/UVOT observation was performed on 2020 May 7 (just 1.4 days after estimated explosion), and provided detections in all bands.

The brightness in the UVOT filters was measured with UVOT-specific tools in  HEAsoft\footnote{\href{https://heasarc.gsfc.nasa.gov/docs/software/heasoft/}{https://heasarc.gsfc.nasa.gov/docs/software/heasoft/} version 6.26.1}. 
Source counts were extracted from the images using a circular aperture with a radius of $3''$. The background was estimated using a significantly larger circular region. The count rates were measured from the images using the \swift\ tool {\tt uvotsource}. They were converted to magnitudes using the UVOT photometric zero points \citep{Breeveld2011a} and the latest calibration files from 2020 September. To remove the host-galaxy contribution, we used observations from before the SN explosion. We measured the flux at the SN site using the same source and background apertures and arithmetically subtracted the host contribution from the SN photometry. All magnitudes were transformed into the AB system using \citet{Breeveld2011a}. These measurements are included in Fig.~\ref{fig:lcs}.

\subsubsection{X-rays}
\label{sec:xrays}  

The field of SN 2020jfo was observed with \swift's onboard X-ray telescope \citep[XRT;][]{Burrows2005a} in photon-counting mode several times between 2020 May 7 and 2021 
January 30. 
\swift\ also observed this field between 2008 and 2016, long before the SN explosion. We analysed all data with the online tools of the UK \swift\ team\footnote{\href{https://www.swift.ac.uk/user_objects/}{https://www.swift.ac.uk/user\_objects/}} 
that use the methods described by \citet{Evans2007a} and \citet{Evans2009a} and the software package 
{HEAsoft}. 

We detect emission at the SN position in the 2008--2016 datasets. 
To recover any SN flux in the later data, we numerically subtracted that baseline flux, 
but this did not result in any significant detections
($<1.4\sigma$).
To convert count-rate limits to flux, we use WebPIMMS\footnote{\href{https://heasarc.gsfc.nasa.gov/cgi-bin/Tools/w3pimms/w3pimms.pl}{https://heasarc.gsfc.nasa.gov/cgi-bin/Tools/w3pimms/w3pimms.pl}}, assume a power law with a photon index
of 2 and a Galactic equivalent neutral-hydrogen column density of $1.58\times10^{20}~{\rm cm}^{-2}$. 
We conclude that any X-ray emission from SN 2020jfo must be fainter than a few $10^{39}\,{\rm erg~s}^{-1}$.

\subsection{Pre-explosion imaging}
\label{sec:preexplosionimaging}

\subsubsection{Progenitor}
The explosion in a nearby Messier galaxy could allow for investigation of the site of the progenitor star. Fortunately, the site of SN 2020jfo was serendipitously imaged prior to explosion with the {\sl Hubble Space Telescope\/} ({\sl HST}) Advanced Camera for Surveys (ACS)/Wide Field Channel (WFC) in bands F435W and F814W (1090~s total exposure time in each band) on 2012
May 24\footnote{GO-12574; PI D.~Leonard} and in F814W (2112~s) on 2019 April 7\footnote{GO-15645; PI D.~Sand}, as well as with the Wide Field Camera 3 (WFC3)/UVIS channel on 2020 March 29\footnote{GO-15654; PI J.~Lee} in F275W (2190~s), F336W (1110~s), F438W (1050~s), F555W (670~s), and F814W (803~s). The SN site is also located in archival {\sl HST\/} Wide-Field Planetary Camera 2 (WFPC2) data, but 
owing to the comparatively inferior spatial resolution and sensitivity of the WFPC2 data, we do not consider them further.

To locate the precise position of the SN in the archival {\sl HST\/} datasets, we obtained high-spatial-resolution images in the $K'$ band with the adaptive optics (AO)-assisted OSIRIS Imager \citep{Larkin2006} on the Keck-I 10~m telescope on 2020 July 7. 
The SN field required laser-guide-star AO and short-turnaround clearance for satellite avoidance.  
There were 10 frames added together, with each coadd consisting of 8 dithered images of duration 14.75~s,
so the total integration time was 1180~s.

\begin{figure*}
\centering
  \begin{minipage}{0.45\textwidth}
\includegraphics[width=\textwidth]{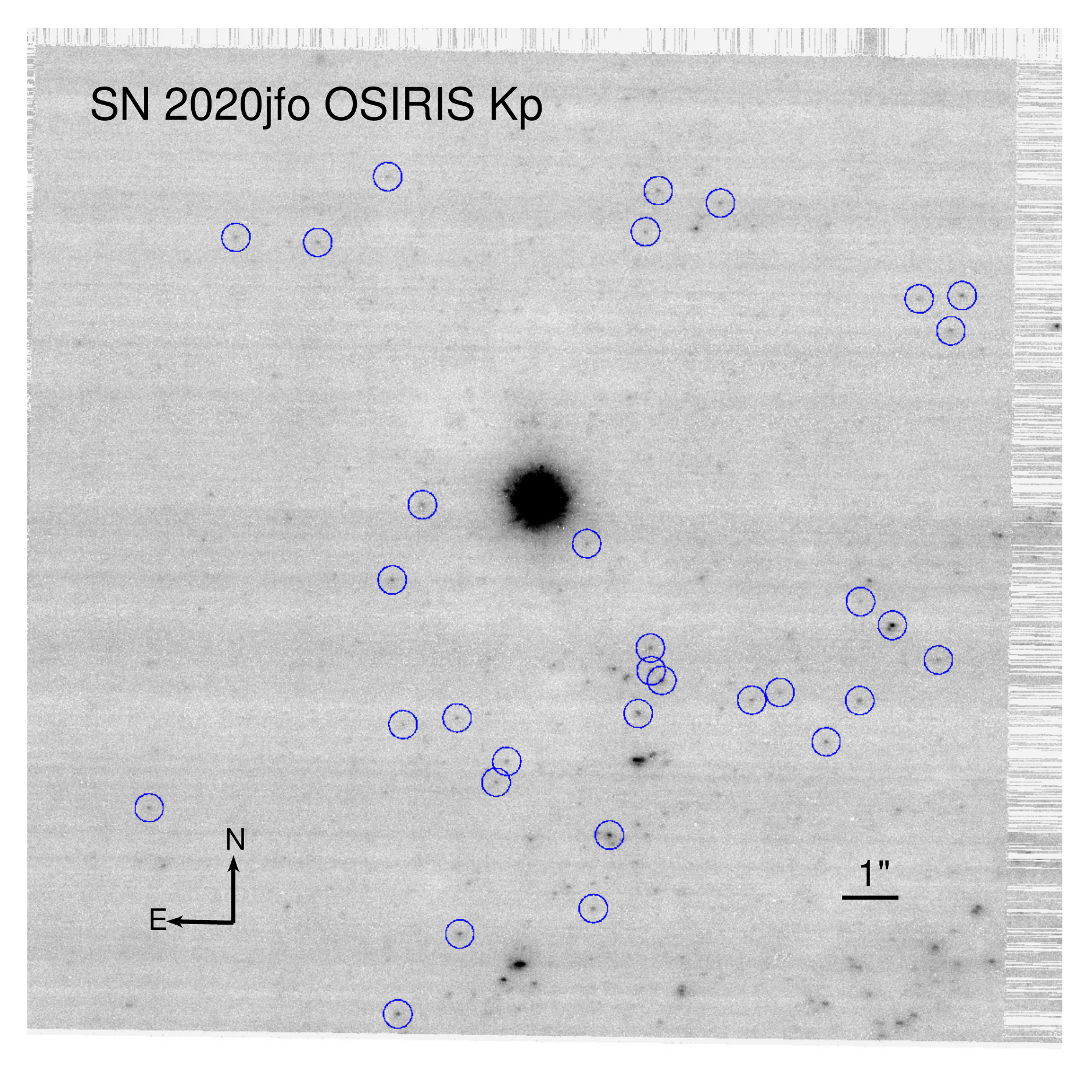}
%\caption{figure caption}
\end{minipage}\hfill
\begin{minipage}{0.45\textwidth}
\includegraphics[width=\textwidth]{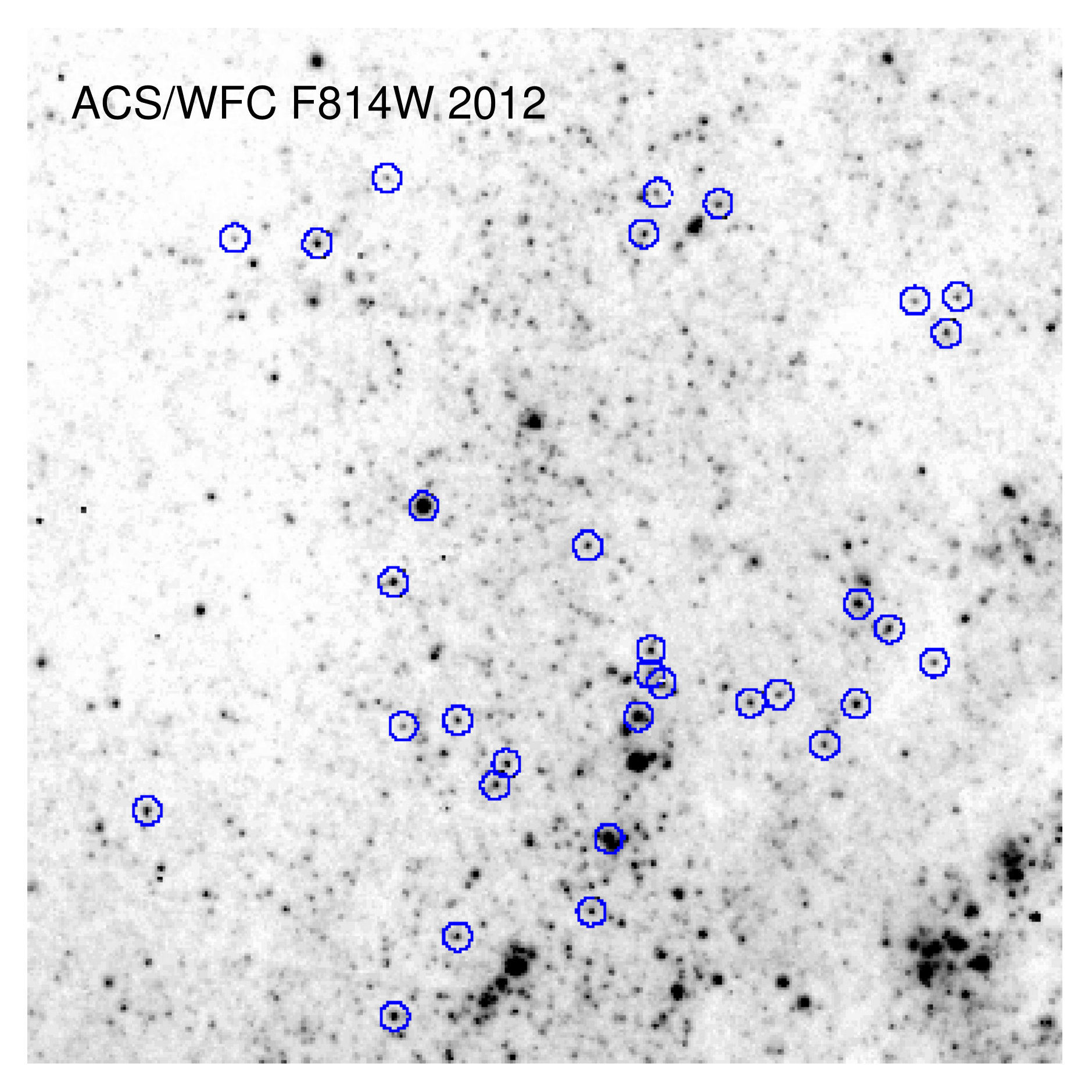}
%\caption{figure caption}
\end{minipage}\par
\vskip\floatsep% normal separation between figures
\begin{minipage}{0.45\textwidth}
\includegraphics[width=\textwidth]{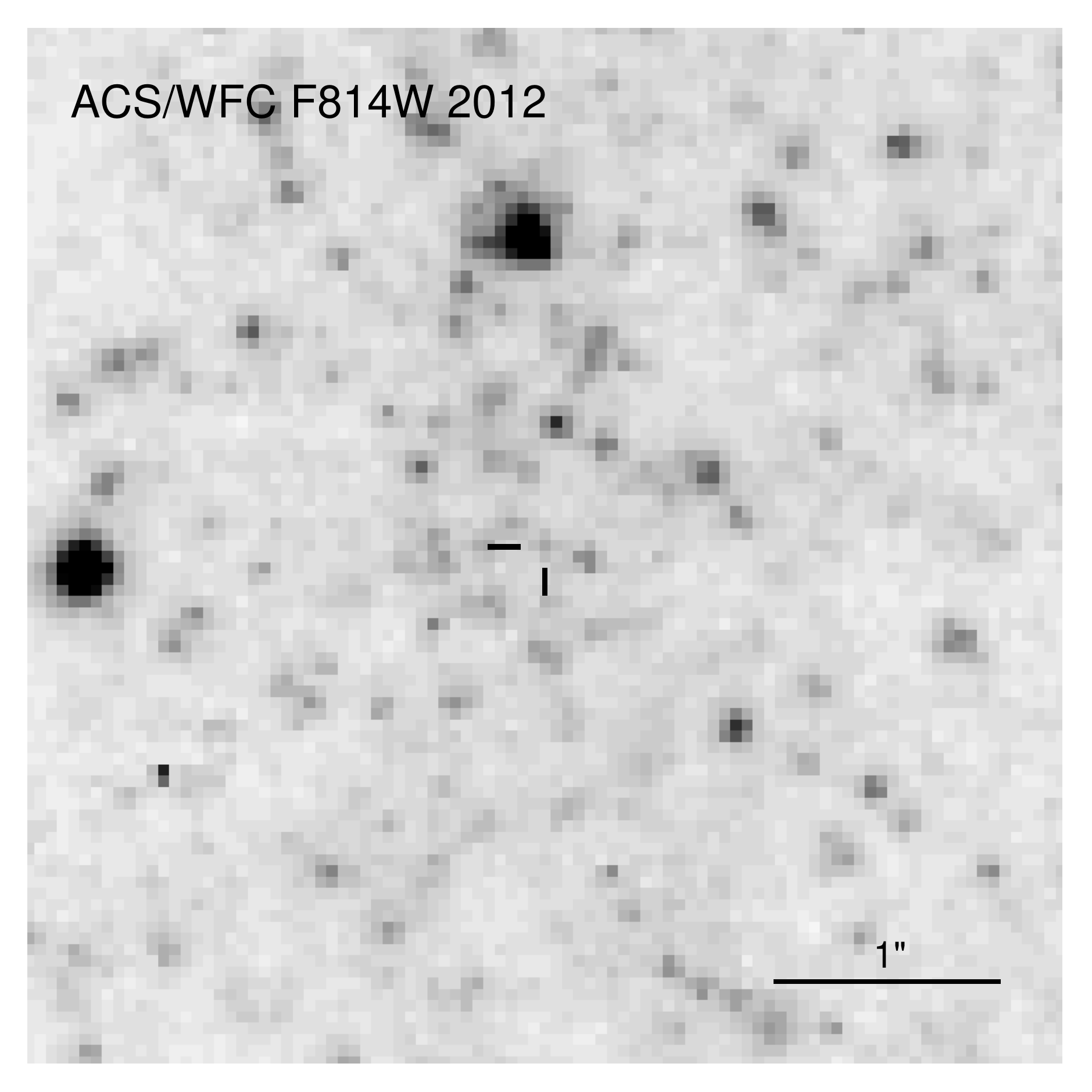}
%\caption{figure caption}
\end{minipage}\hfill
\begin{minipage}{0.45\textwidth}
\includegraphics[width=\textwidth]{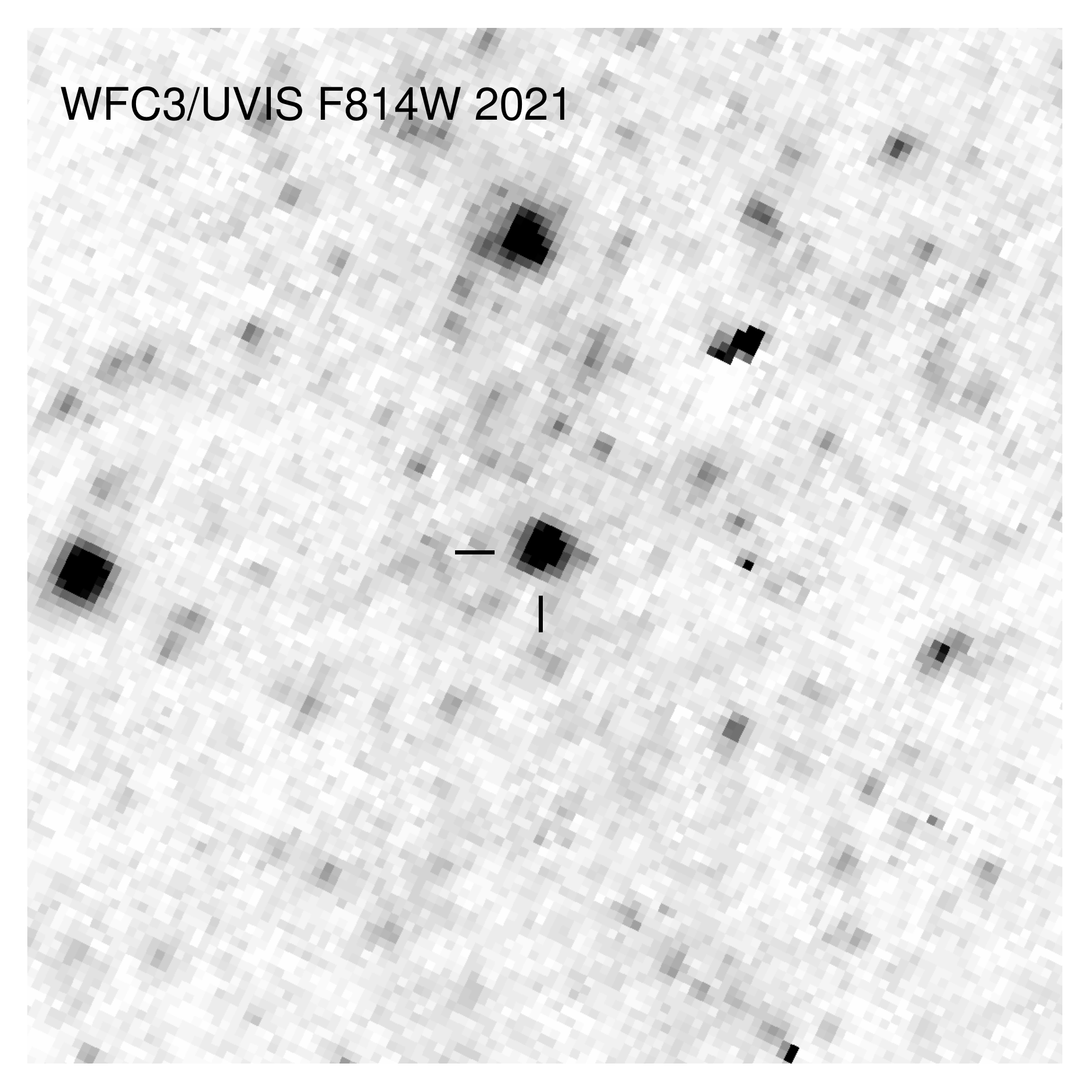}
%\caption{figure caption}
\end{minipage}\par
    \caption{{\it Top left}: Adaptive optics (AO) image of SN 2020jfo obtained with the OSIRIS imager on the Keck-I 10~m telescope in the $K'$ band on 2020 July 7. {\it Top right}: Portion of an archival {\sl HST\/} ACS/WFC image mosaic obtained in F814W on 2012 May 24, which contains the SN site, shown at the same scale and orientation as the top-left panel. In both top panels, the 32 stars in common between the two images and used in the astrometric registration are circled. {\it Bottom left}: A zoom-in on the SN site in the {\sl HST\/} image, with the SN progenitor candidate indicated by tick marks. 
    {\it Bottom right}: {  Zoom-in on the SN site after the explosion, in a {\sl HST\/} WFC3/UVIS F814W image from 2021 July 28. The SN is clearly identified, confirming the progenitor candidate identification based on the AO image. North is up and east is to the left in all panels.}
    }
    \label{fig:hst}
\end{figure*}

We astrometrically registered the Keck AO image to the 2012 ACS/WFC F814W image using 32 stars in common between the two image datasets, employing the task {\tt geomap} in PyRAF with parameter ``calctype" set to ``double."
The geometric distortion for the OSIRIS AO imager is quite small (${<}5$ milliarcsec), so a distortion correction is not strictly required\footnote{J.~Lu, private communication}.
We were able to achieve a satisfactory solution to a precision of 0.15 WFC pixel (7.5 milliarcsec). Figure~\ref{fig:hst} shows the Keck and ACS images in a broader view, as well as a zoom-in on the SN site. 
With regards to a version of the 2012 F814W image mosaic available from the Hubble Legacy Archive\footnote{http://hla.stsci.edu/}, from the position of the SN in the OSIRIS image, and referencing our astrometric solution, we expect the precise SN location at pixel (2110.05, 3062.02). In the image mosaic a faint object can be seen with a centroid position of (2110.06, 3061.93). The difference between these two pixel values is within the astrometric uncertainty, and we therefore consider the object to very likely 
be the candidate for the SN progenitor.

In order to determine the brightness of this object from the archival {\sl HST\/} data, we extracted photometry from the individual 
frames for both the ACS and WFC3 observations with Dolphot \citep{Dolphin2016}. We used recommended parameters for Dolphot appropriate for a crowded extragalactic environment\footnote{FitSky=3 and RAper=8}. The object is only detected in F814W. We measured F814W = $25.02 \pm 0.07$ mag in 2021, and $25.56 \pm 0.09$ and $25.08 \pm 0.13$ mag in 2019 and 2020, respectively. 
Upper limits were placed in the other bands; these are all at the
5$\sigma$ level. The detection limit in 2012 in F435W is 26.8. The limits in 2020 are 25.1, 25.2, 26.5, and 26.8 mag in F275W, F336W, F438W, and F555W, respectively. The lack of detection of the star in bands bluer than F814W indicates that this is a cool, red star.

The detected star, given the assumed distance modulus and accounting only for MW extinction, has absolute magnitudes of $M_{\rm F814W} \approx -5.8$, $M_{\rm F275W} > -5.8$, $M_{\rm F336W} > -5.7$, $M_{\rm F438W} > -4.4$, and $M_{\rm F555W} > -4.1$. For red supergiants (RSGs) near the terminus of their evolutionary tracks, from the BPASS solar-metallicity single-star models \citep{Stanway2018} at 10--15 \Msun, we would expect $M_{\rm F814W} \approx -7$ to $-8$ mag. This implies that the progenitor may have been either more compact or further extinguished by $A_{\rm F814W} \gtrsim 1$ mag, potentially from circumstellar dust. 

{  We also obtained image data in F555W and F814W on 2021 July 28 with the {\sl HST\/} 
WFC3/UVIS\footnote{GO-16179 PI: A.~Filippenko}. The SN is still prominent in the images (Fig.~\ref{fig:hst}), and we confirm, with an astrometric 1$\sigma$ uncertainty of 0.15 UVIS pixel (6 milliarcsec), the identification of the progenitor candidate, based on the AO image.}

\subsubsection{Precursor study}

\begin{figure*}
\centering
     \includegraphics[width=0.8\textwidth]{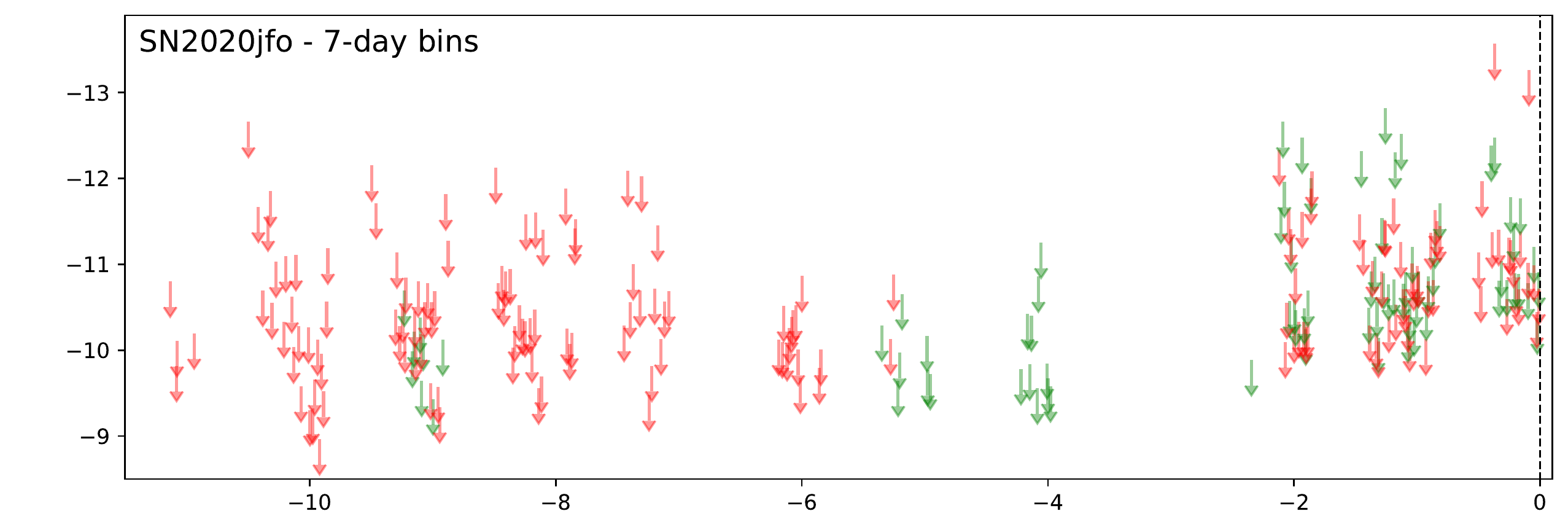}\\
     \includegraphics[width=0.8\textwidth]{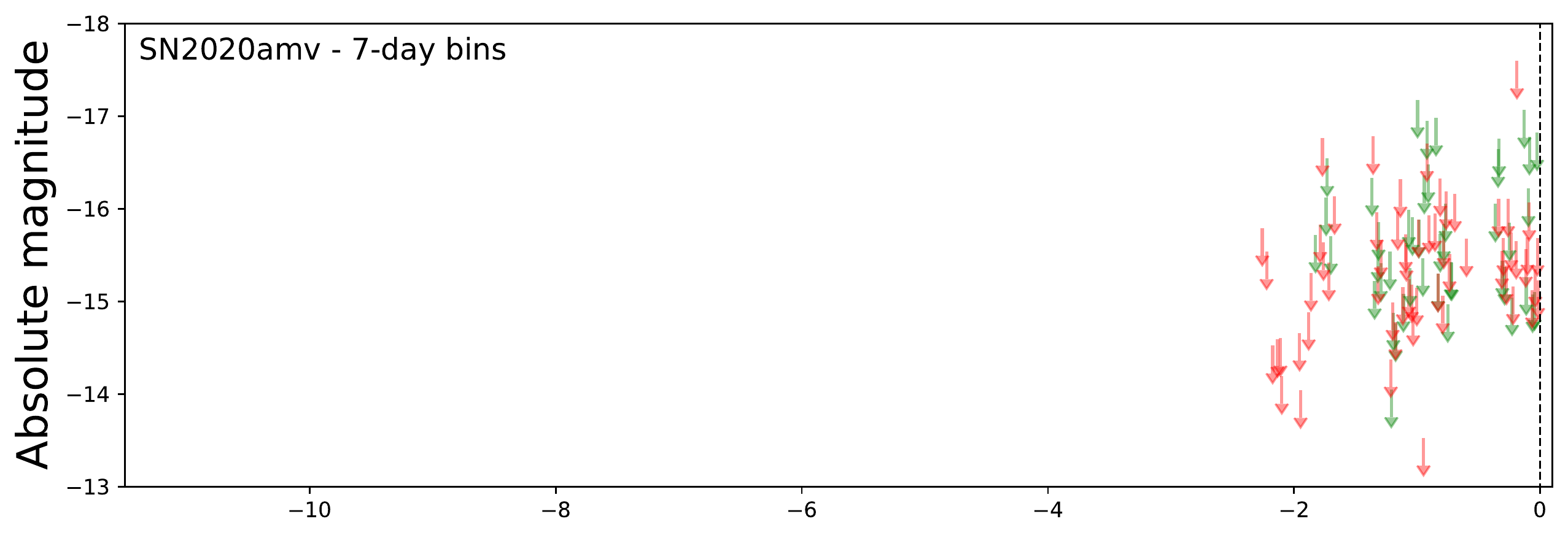}\\
     \includegraphics[width=0.8\textwidth]{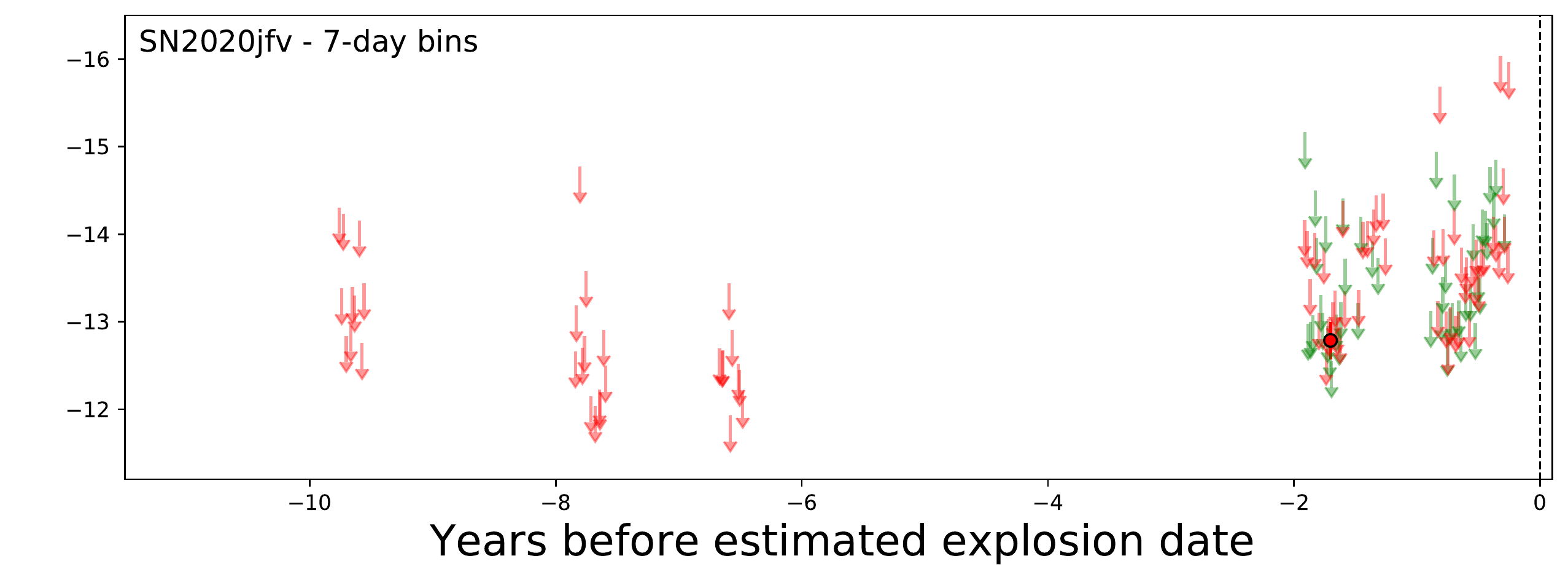}
    \caption{Pre-explosion light curves from both (i)PTF and ZTF of the three SNe in 7-day bins. No firm detections were obtained in the 11~yr prior to explosion, and the limits are discussed in the text. 
    }
    \label{fig:precursor_search}
\end{figure*}

The ZTF survey first started to monitor the position where SN 2020jfo exploded more than 800 days (2.4~yr) before the explosion date. We obtained in total 300 pre-explosion observations during 158 different nights. No precursors were detected when searching the unbinned or binned data following the methods described by \citet{strotjohann2021}. The median limiting magnitude is $-11$ in the $r$ band and brighter precursors can be excluded in 31 weeks or 25\% of the time, while outbursts as bright as magnitude $-12$ can be ruled out 49\% of the time within the last 2.4~yr before the SN explosion.
The SN location was also regularly observed by the Palomar Transient Factory (PTF) and the intermediate PTF (iPTF) surveys; 988 observations were obtained spanning 11.1 to 4~yr before the SN explosion. We can rule out week-long precursors that are brighter than magnitude $-11$ in the Mould $R$ band in 102 weeks or 28\% of the time. The upper panel of Fig.~\ref{fig:precursor_search} shows the limiting magnitudes for week-long bins.

A similar precursor search for SN 2020amv also did not reveal any pre-explosion activity, as shown in the middle panel of Fig.~\ref{fig:precursor_search}, but the limits are less constraining owing to the larger SN distance. The median limiting magnitude is $-15.7$ in the $r$ band and such a bright outburst can be excluded in 30 weeks or 25\% of the time. The detected flash-spectroscopy features in the early-time spectra of SN 2020amv indicate that the progenitor star lost material shortly before the explosion, but apparently this mass-loss event was not associated with any optical outburst that was bright enough to be detected in this search. This is consistent with \citet{strotjohann2021}, who observed no pre-explosion outbursts prior to 20 SNe with flash-spectroscopy features, including SN 2020amv, even though several of them were located at small redshifts of $z < 0.02$. 
According to \citet{strotjohann2021}, this indicates that these flash-spectroscopy SNe likely have fewer or fainter outbursts than Type IIn SNe, but their sample was too small to constrain the precursor rate further. 
SN 2020amv reveals, as described below, both early flash features, but also later evidence for CSM interaction.

For SN 2020jfv, a single bin surpasses the formal $5\sigma$ threshold of the precursor search. The detection occurs 1.8~yr before the estimated explosion date in the $r$ band and is seen when combining data in 7-day or 30-day bins. However, a more detailed inspection shows that this is likely a false detection. No point source is seen when coadding the $3$ difference images in the bin, and the $g$-band images in the same nights yield a significantly deeper limiting magnitude.
The median limiting magnitude is close to $-14$ in the $r$ band and we can exclude such a bright precursor in 29 weeks during the final 2~yr before the SN explosion (28\% of the time), assuming that the mentioned single detected flux excess is not real. We also analyzed 158 PTF/iPTF observations and can exclude precursors brighter than magnitude $-14$ in 31 weeks.

\subsection{Optical spectroscopy}
\label{sec:opticalspectra}

Follow-up spectroscopy of SN 2020jfo was primarily conducted with robotic telescopes, most of them with the SEDM mounted
on the P60, but also with SPRAT on the LT.
Further spectra were obtained with the NOT using ALFOSC. This included the above-mentioned classification spectrum for SN 2020jfo, but also later nebular-phase spectra. We also obtained spectra with the Lick 3~m Shane telescope equipped with the Kast spectrograph.
The full log of spectra is provided in Table~\ref{tab:spec}.
As can be seen in the table, P60+SEDM was also instrumental in providing spectra for the other two SNe. 
For the three SNe, {  42} spectra in total where obtained. 
Additional spectra in this paper come from Gemini-North equipped with GMOS, the Palomar P200 telescope and DBSP  \citep{Oke1982}, as well as deep nebular spectra taken with the Keck-I telescope using the Low Resolution Imaging Spectrograph (LRIS; \citealp{Oke1995}).

The \texttt{LPipe} reduction
pipeline \citep{perleyspec}
was used to process the LRIS data. SEDM spectra were reduced using the pipeline described by
\citet{rigault}, and the
spectra from La Palma were reduced using standard pipelines and procedures for each telescope and instrument.
The ALFOSC spectra were often reduced using {\tt PypeIt} \citep{pypelt}. 
All spectral data and corresponding information will be made available via WISeREP\footnote{\href{https://wiserep.weizmann.ac.il}{https://wiserep.weizmann.ac.il}} \citep{Yaron:2012aa}. All spectra have been calibrated on an absolute scale using contemporaneous (or interpolated) photometry.

\subsubsection{Spectropolarimetry}\label{sect:specpol}

We obtained two epochs of spectropolarimetry of SN 2020jfo on the nights of 2020 May 25 and May 29 
(19.7 and 23.7 rest-frame days past explosion)
using the polarimetry mode of the Kast spectrograph on the Lick 3~m Shane telescope. On each night, low-polarization and high-polarization standard stars were observed to calibrate the data. Observations and data reduction were carried out as in Patra et al. (in prep.).

Linear polarization is calculated from the Stokes $Q$ and $U$ parameters as $P = \sqrt{Q^{2} + U^{2}}$, and the polarization position angle (PA) on the sky is defined as $\theta = 1/2 \arctan(U/Q)$. 
$P$ 
is a positive-definitive quantity and therefore overestimated when the signal-to-noise ratio (S/N) is low. We debias the polarization as \begin{equation}
P_{db} = P - \left (\frac{\sigma^{2}_{P}}{P} \times h(P - \sigma_{P}) \right) ~ \mbox{and} ~~ \theta_{db} = \theta,
\end{equation}
where $\sigma_{P}$ is the $1\sigma$ uncertainty in $P$ and $h$ is the Heaviside step function.
 
%\paragraph{Instrumental and interstellar polarization}
We measured the average Stokes $Q$ and $U$ of the low-polarization star HD 110897 to $<0.05 \%$, demonstrating the low instrumental polarization. We also find low interstellar polarization (ISP) in the direction of SN 2020jfo. 
\cite{1975ApJ...196..261S}
showed that an upper limit on ISP owing to dichroic extinction by dust grains can be derived as $9 \times E(B-V)$. In the direction of SN 2020jfo, the estimated $E(B-V) = 0.02$ mag 
implies that ISP $< 0.18$\%. We confirm that the Galactic ISP is low by measuring the polarization of an ISP "probe-star"\footnote{We observed the star Gaia ID 3894181087039808128.}, an unpolarized star close to the line of sight to SN 2020jfo. We found the polarization of the probe star to be $<0.15$\%. Furthermore, the emission peak of the H$\alpha$ feature of a SN is expected to be depolarized due to contamination by unpolarized flux diluting the underlying polarized continuum flux. The minimum observed polarization in the H$\alpha$ feature also constrains the ISP to be $< 0.2$\%. Taken together, the three lines of evidence all show that the ISP in the direction of SN 2020jfo is low. 

The continuum polarization, which reflects the global ejecta asymmetry, is 0.4-0.7\% over the two epochs, typical of SNe II while on the plateau 
\citep[see][and references therein]{2008ARA&A..46..433W}.

\begin{figure*}
\centering
     \includegraphics[width=\textwidth]{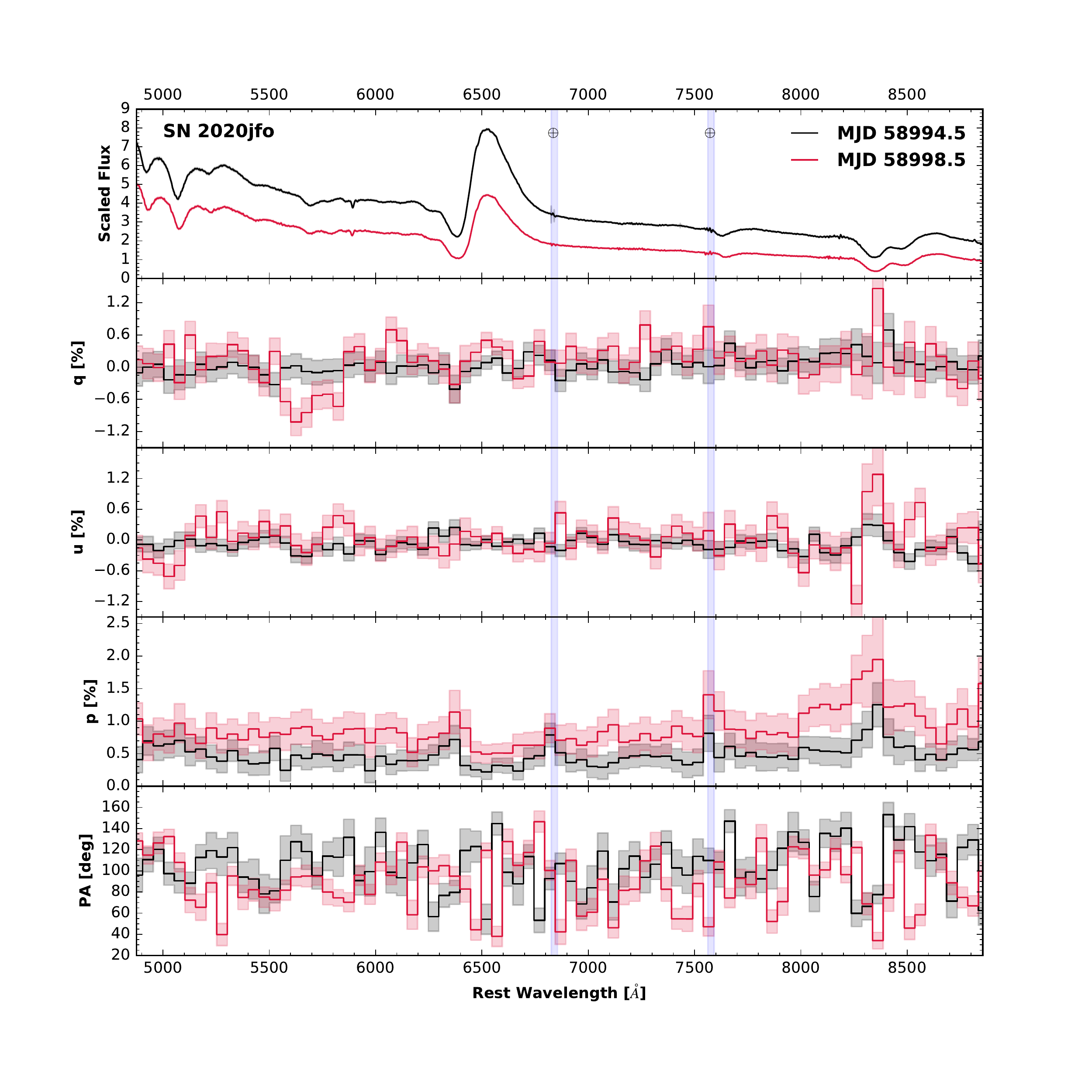}
    \caption{
    Spectropolarimetry of SN 2020jfo on 2020 May 25 and May 29. The panels (from top to bottom) show total flux, Stokes $Q$, Stokes $U$, debiased polarization, and the position angle. The vertical bands represent the regions of potential telluric overcorrection. The data, except for total flux, are binned to 50~\AA\ to improve the S/N. 
    }
    \label{fig:specpol}
\end{figure*}

The polarization position angle hovers around a mean of
$\sim100\degr$
over the two epochs. The lack of significant change in PA could imply a global axis of symmetry, 
although the data do not have sufficiently long temporal coverage to confirm this. 
The 
\ion{Ca}{II}
NIR feature is significantly polarized, with levels exceeding 1\% at both epochs
(see Fig. \ref{fig:specpol}). 
The high line polarization suggests that Ca is not uniformly distributed within the ejecta and likely exists in clumps. The \ion{Ca}{II} line polarization also shows velocity-dependent variation, with the 
high-velocity feature more polarized than the 
normal-velocity feature.

\section{Results and Discussion}\label{sec:discussion}

\subsection{Light curves}

The LCs of our three SNe are displayed in Fig.~\ref{fig:lcs}.  
We first focus on the LC of SN 2020jfo.

\subsubsection{SN 2020jfo}\label{sect:2020jfolc}

As mentioned above, 
we have fairly good constraints on the explosion epoch with an uncertainty of 
$\pm2$ days simply based on the last nondetection.
Therefore, we can also measure the rise time with some precision.
We used a Gaussian Processing (GP)
algorithm\footnote{\href{https://george.readthedocs.io}{https://george.readthedocs.io} {  with a Matern32 kernel.}}
to interpolate the photometric data and measure the rise time and the peak magnitudes. 
The results are provided in Table~\ref{table:SNLCproperties}. The SN rises to maximum brightness in slightly less than 5 days in both $g$ and $r$. The peak magnitude of $m_r^{\rm{peak}} = 14.4$ made SN 2020jfo one of the brighter
CC~SNe during 2020.  

After the initial rise follows a plateau phase in the $r$ band of $\sim60$+ days, which establish SN 2020jfo as a Type IIP SN.  The $i$ band is well sampled the first 40+ days, when it follows the plateau, whereas the $g$ band declines faster. The LCs cover the first 60 days well, whereafter the SN position was too close to the Sun in the sky.  It was recovered after solar conjunction in $g$ and $r$, declining linearly up to $\sim350$ days.
The forced-photometry ATLAS LCs confirm the Type IIP classification and the general shape of the LC. In fact, the fall off the plateau is best seen in the $o$ band, which indicates a plateau length of  
$64 \pm 3$ 
days\footnote{As done by \cite{Anderson_2014}, we fit $o$-band data using a $\chi^2$ minimising procedure with a composite function of a Gaussian, Fermi-Dirac, and a straight line, following \cite{Olivares_E__2010}.}.
In comparison with the large SN II sample of 
\cite{Anderson_2014}, this is actually one of the shortest plateau lengths.
We follow the SN up to a year after explosion.

The colour evolution in $g-r$ for SN 2020jfo 
and also for the Type II SN 2020amv initially become redder for the first 60 days, while SN 2020jfo was on the plateau. At about 200 days past explosion all our three SNe display
$g-r\approx1.0$ mag. 
SN 2013ej \citep{valenti2013ej,Fang2016MNRAS.461.2003Y}, which is
known to experience little host-galaxy extinction, show a similar
initial reddening.
Since our three SNe are bluer than SN 2013ej, this at least conforms with our omission of extra host extinction corrections.

\begin{figure*}
\centering
    \includegraphics[width=0.8\textwidth]{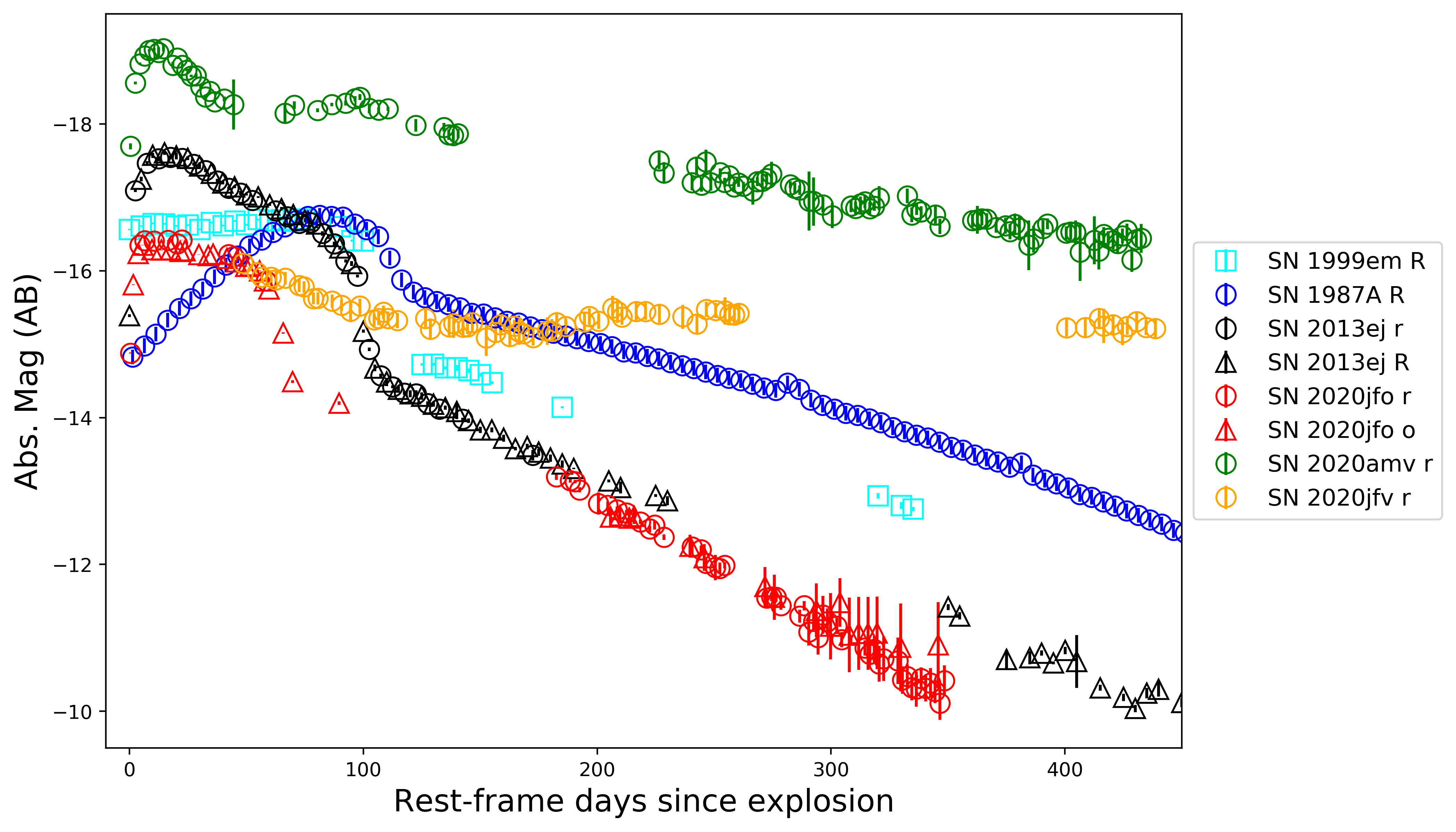}
    \caption{Light curves in absolute magnitudes
    ($M_{{r}}$) 
    for our three SNe. This accounts for distance modulus and MW extinction as discussed in the text, but no additional corrections for host extinction. The comparison SNe are introduced in the text. 
    The photometry has been binned to nightly averages.
    }
    \label{fig:lcabs}
\end{figure*}

In Fig.~\ref{fig:lcabs} we show the LCs in
absolute magnitudes ($M_r$) together with the LCs of a few other SNe~II. 
The magnitudes in Fig.~\ref{fig:lcabs} are in the AB 
system\footnote{The Vega/AB magnitude conversion follows \cite{Blanton2007}.}
and have been corrected for distance modulus, MW extinction, and host extinction if any, and are plotted
versus rest-frame days past estimated explosion epoch. SN 2020jfo reached a peak magnitude quite similar to that of the canonical Type IIP SN 1999em\footnote{
We used $E(B-V) = 0.035$ mag and a distance of 7.5 Mpc from \cite{Hamuy2001} for SN 1999em, data from \cite{Faran2014MNRAS.442..844F}.
}, 
but the plateau phase is significantly shorter. At nebular phases SN 2020jfo is fainter than SN 1999em, but instead follows the same decline rate and tail luminosity as the Type II
SN 2013ej\footnote{ 
We used $E(B-V) = 0.060$ mag and a distance of 9.1 Mpc from \citet{valenti2013ej,Fang2016MNRAS.461.2003Y} for SN 2013ej.}.

\begin{figure*}
\centering
     \includegraphics[width=0.9\textwidth]{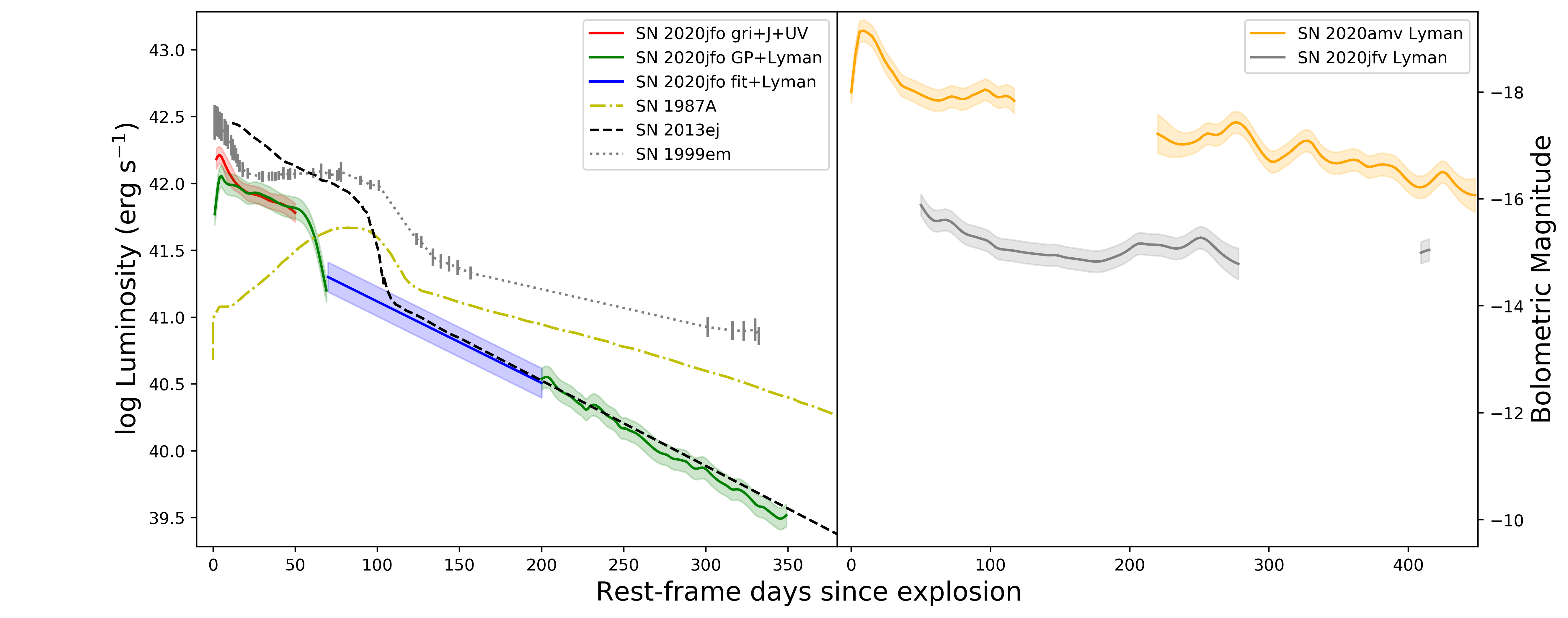}
    \caption{Bolometric luminosities for our SNe. Left: SN 2020jfo with some comparison objects (see text). The green solid lines with shaded regions are estimated with the \cite{2016MNRAS.457..328L} method and is a Gaussian Process fit to that LC. The  
    blue region is a linear interpolation for the region with less data. The red fit on the plateau is for a diluted BB fit to our UV through NIR data, and matches the Lyman et al. method very well {  on the later part of the plateau}. Right: SNe 2020amv and 2020jfv. These were derived using the Lyman et al. method, which is rather approximate given the unusual nature of these SNe.}
    \label{fig:lum}
\end{figure*}

In order to estimate the total radiative output, we also attempted to construct bolometric LCs.
For SN 2020jfo, we adopted two approaches: we use a black-body (BB) function fitted to the GP interpolated fluxes, and we also employ an analytic bolometric correction (BC) that was constructed by \cite{2016MNRAS.457..328L} from a sample of SNe~II.
On the LC plateau, where we have coverage from the NIR to the UV, we
can construct spectral energy distributions (SEDs) and fit to a
diluted BB function. Integrating that BB function provides the
bolometric luminosity.
 Comparing this with the prescription from \cite{2016MNRAS.457..328L}, we find that the bolometric LC on the plateau agrees very well with our BB estimate 
{  (the ratio is $0.97 \pm 0.05$ on the late plateau).} 
 Therefore, for the part of the LC where we only have optical data, we follow the \citet{2016MNRAS.457..328L} method. {  The only difference between these methods is for the very early phases, where our UV data imply slightly higher luminosities; this is illustrated in Fig.~\ref{fig:lum}.} The bolometric LC is also provided on WISEREP.
 
Using this, we can estimate a 
maximum bolometric luminosity for SN 2020jfo 
{  of $L_{\rm{bol}} = 1.63 \times 10^{42}$ erg s$^{-1}$}
at 4 rest-frame days
and a total radiated energy over the first 350 rest-frame days 
{  of $E_{\rm{rad}} = 5.74 \times 10^{48}$ erg.} 
The radioactive $^{56}$Ni mass ejected in the explosion can be inferred by measuring the luminosity tail, 
which is powered by the decay of radioactive $^{56}$Co.
Using 
$L = 1.45 \times 10^{43}$ exp$(-t/{\tau_{\rm Co}}) ({M_{\rm Ni}}/{{\rm M}_{\sun}}) $ erg s$^{-1}$ from \cite{Nadyozhin} implies that we would require 
$0.024 \pm 0.002$ \Msun\ of $^{56}$Ni to account for the luminosity (70--100 rest-frame days; see further 
Sect. \ref{sect:MCMC}).

\subsubsection{Comparisons with SNe 2020amv and 2020jfv}

We here discuss the LCs of the other two SNe presented in this paper, 
although not at the same level of detail as for SN 2020jfo.
SN 2020amv was photometrically monitored with P48 for more than a year, with a few data points also provided by LT, P60, and NOT. The LC is displayed in the middle panel of Fig.~\ref{fig:lcs}. The explosion date is well constrained and the SN rose in a bit more than two weeks ($r$ band; Table~\ref{table:SNLCproperties}) to a Gaussian-shaped LC peak, where both the rise and the fall are somewhat faster in $g$ than in $r$, with
${\Delta}{m_{15}^{r}} = 0.45 \pm 0.02 $ mag and 
${\Delta}{m_{15}^{g}} = 0.52 \pm 0.01 $ mag.
After this initial phase, the LC is rejuvenated, as the $r$-band brightness gently rises again some 60 days past peak. Overall, it is a very long-lived SN which we followed for more than 450 days.
The absolute-magnitude LC (Fig.~\ref{fig:lcabs}) demonstrates that SN 2020amv was very luminous, $M^{\rm{peak}}_{g} = -19.2$ mag, the initial peak resembling that of a Type I SN, but the remaining bumpy, bright, and long-lived LC reveals that the (late-time) power source must be something in addition to radioactive decay. In terms of CC SNe, such LCs are expected to be powered by CSM interaction \citep[e.g.,][]{Nyholm2017,Nyholm2020}, but we note that superluminous SNe, even of Type I, sometimes display long-lived bumpy LCs, which are not always easily explained in terms of a central engine (e.g., the large SLSN-I sample from ZTF; Chen et al. 2021, in prep.).

SN 2020jfv was found while declining and we do not have good constraints for the date of explosion or the early peak. The right-hand panel of Fig.~\ref{fig:lcs} therefore only shows a declining and eventually flattening LC. The unusual aspect is that the $r$-band LC, after declining a full magnitude over the first $\sim 100$ days, rises by $\sim0.3$ mag in the next 100 days, before the target was lost in the Sun's glare. {  After solar conjunction, we recovered the LC in $gri$ at virtually the same magnitudes as before the gap.}

\subsubsection{Light-curve modeling for SN 2020jfo}\label{sect:MCMC}

In order to estimate progenitor and explosion parameters for SN 2020jfo from the bolometric LC, we make use of the semi-analytic Monte Carlo code that was recently presented by \citet{Jager2020}, 
as used for the low-luminosity Type IIP 
SN~2020cxd {  \citep{yang2021arXiv210713439Y}}.
After marginalisation, our fit provides estimates with confidence intervals (2$\sigma$) for each of the parameters:
SN\,2020jfo has 
$M_{\rm ej}=5.16_{-2.00}^{+0.28}$~\Msun,  
$E_{\rm kin}=2.04_{-1.02}^{+0.57}\times10^{51}$ erg, and
$v_{\rm exp}=8.14_{-1.32}^{+0.54} \times10^{3}$~km~s$^{-1}$
for the ejecta mass, kinetic energy, and expansion velocity
(respectively). The nickel mass was simultaneously estimated as $0.029\pm0.014$ \Msun.
The mass of radioactive nickel is thus similar to that estimated for SN 2013ej \citep[0.023 \Msun;][]{Fang2016MNRAS.461.2003Y}, as expected from the similar absolute magnitudes (Fig.~\ref{fig:lcabs}). It also matches the estimate from Sect.~\ref{sect:2020jfolc}. 
Moreover, the LC slope is similar for these two SNe, and \cite{Fang2016MNRAS.461.2003Y} interpreted this as being due to gamma-ray escape. The same seems to apply here. In fact, we can fit for a gamma-leakage LC \citep{sollerman1998ApJ...493..933S} with
the flux declining as
$     e^{(-t/111.3)} \times (1-0.965\times e^{-(t_0 / t)^2}),$
where $t$ is the time in days and $t_0$ is the epoch when the optical depth to the gamma rays is unity.
This epoch is also related to the ejecta mass \citep[][their Eq.~5]{clochiatti1997ApJ...491..375C}, and our best-fit values
$M_{\rm {Ni}} = 0.025$~\Msun\ and $t_0 = 166$ days correspond to $M_{\rm ej}\approx5-6$ \Msun, for (1--1.7) $\times10^{51}$ erg  of kinetic energy.
We note that the estimated ejecta mass is low, in agreement with the results from the Monte Carlo fit to the short plateau.

In this respect, the short plateau of SN 2020jfo is similar to those discussed by
{ 
\cite{hiramatsu2021ApJ...913...55H}.
}
They presented three Type IIP SNe having plateau lengths of only 50--70 days, arguing that this was due to low ejecta mass. They further suggested that their SNe originated from massive progenitors with normal to large amounts of radioactive nickel, and that the required large mass loss was also likely the cause for the bright early luminosity. For SN 2020jfo we have a normal initial luminosity and a modest amount of radioactive nickel. There are no signs of CSM interaction from either the LC or the spectra. In addition, we have evidence that the initial mass was not very large (Sect.~\ref{sect:nebularoxygenlines}). 
SN 2020jfo is therefore somewhat different from the discussion of \cite{hiramatsu2021ApJ...913...55H} in the sense that the plateau is not generally powered by the radioactivity.

\subsection{Spectroscopy}

\begin{figure*}
\centering
     \includegraphics[width=0.8\textwidth]{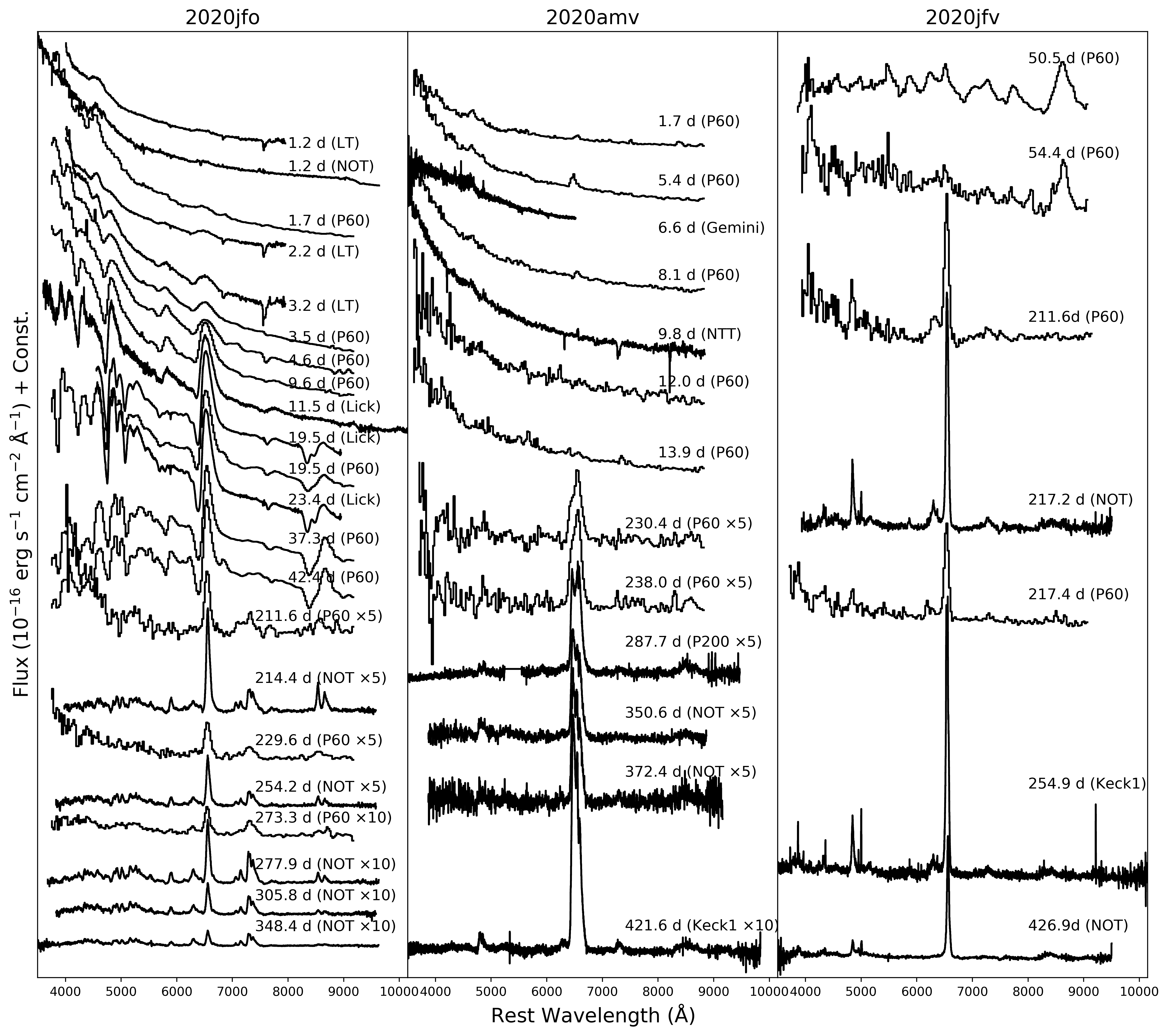}
    \caption{Sequence of spectra of SNe 2020jfo, 2020amv, and  2020jfv. 
    The epochs and scale factors of the spectra are also provided.
}
    \label{fig:spectra}
\end{figure*}

The spectroscopic sequence for SN 2020jfo, as provided in Table~\ref{tab:spec}, is displayed in Fig.~\ref{fig:spectra}.
The photospheric phase is well covered by the robotic telescopes, and although the resolution of the SEDM spectra is low, the overall spectral evolution as the photosphere expands and cools is nicely covered. The early classification spectra show shallow Balmer lines and a \ion{He}{II} feature at 15,000 km s$^{-1}$. No signatures of CSM interaction are present in the spectra. A total of 22 spectra are presented for SN 2020jfo, covering phases from 2 to 350 days past explosion.  The photospheric velocities, as estimated from the P~Cygni Balmer lines, is $\sim9000$~km~s$^{-1}$ at the early plateau and $\sim7000$~km~s$^{-1}$ toward the end of the plateau (44 days past explosion).

\subsubsection{SN 2020amv}

 Thirteen spectra were obtained of SN 2020amv (Table~\ref{tab:spec}). 
 The first spectrum was acquired with the SEDM about 1.7 days from the estimated explosion date. 
 Despite the low resolution, 
 several narrow emission lines
 are identified (e.g., H${\alpha}$, H${\beta}$),
 including those of highly ionised species 
 (\ion{He}{II} $\lambda$4686) which correspond to flash-ionisation lines \citep{Bruch_2021}. 
 These features disappear within 13 days from the explosion epoch. 
 This is thus one of the more long-lived flash features in the ZTF sample (Bruch et al, in prep.).
 Such transient emission lines emerge from the early interaction of the shock-breakout radiation with a nearby (typically $\lesssim10^{15}$cm; e.g., \citealt{Yaron}) CSM. 
 Two higher-resolution spectra were obtained at 6.6 and 9.8 days.
 The narrow H$\beta$ emission line in the former indicates a relatively slow velocity of 
$\lesssim 335$ km~s$^{-1}$ 
(full width at half-maximum intensity; FWHM),
consistent with a wind velocity of the CSM. 
The following spectra show mainly a blue continuum. After the rebrightening of the light curve was recognised,  additional spectra were obtained with SEDM 230 days past explosion. These showed a strong H$\alpha$ line, with very high velocities.
This triggered us to obtain a higher-resolution spectrum with the P200 (Table~\ref{tab:spec}), revealing a broad, boxy line profile having several emission peaks. This complex line profile slowly evolves in later spectra, and the nebular lines are further explored in
Sect.~\ref{sect:nebularlines}.

SN 2020amv can in some sense also be seen as a transforming SN (as we argue for SN 2020jfv below). It seems that CSM interaction is the driving force in powering SN 2020amv, but that the evidence for this is manifested in different ways throughout the SN evolution. The early flash spectroscopy provides evidence for dense CSM close to the exploding star,
although the early LC is similar to Gaussian-shaped LCs of other types of stripped-envelope SNe. Other studies have found that such LCs might still be consistent with radioactive powering, as in SN 2018ijp {  \citep{Tartaglia2021} 
},
or SN 2020eyj (Kool et al., in prep.), 
where CSM interaction came in later to power the prolonged LCs. SN 2020amv peaks at $-19.24$ mag in $g$, similar to the peak luminosity of SNe Ia. The CSM evidence in terms of LC evolution is instead obvious after about 50 days, when the second peak anticipates the long-lived nature of the overall LC. Here we lack spectroscopy, but when the spectroscopic campaign resumed it revealed nebular box-shaped emission lines (Sect.~\ref{sect:nebularlines}) that again are a telltale signature for CSM interaction --- this time the signature unfolds
a cold dense shell likely caused by the reverse shock produced by ejecta running into the CSM. These are all different manifestations of CSM interaction that together unveil the mass-loss history of the exploded star.

\subsubsection{SN 2020jfv}

SN 2020jfv was first spectroscopically observed with the SEDM on P60, and based on this spectrum classified as a Type IIb SN
\citep{2020TNSCR1948....1D}.
Helium lines at a velocity of 
7000~km~s$^{-1}$ are detected. 
Six months later, after recognising that the target was rebrightening again, we obtained another SEDM spectrum that seemed to be dominated by bright, narrow Balmer emission lines, typical for CSM-driven SNe~IIn.  
We also acquired some spectra with larger telescopes, namely with the NOT and Keck (Fig.~\ref{fig:spectra}). These show a Type II SN spectrum containing lines from elements such as Mg, Ca, and O (as are typically seen in CC SNe), but which is very much dominated by the intermediate-width Balmer lines. We discuss the nebular spectra in the next sections.

\subsection{Modelling the oxygen mass of SN 2020jfo}\label{sect:nebularoxygenlines}

\begin{figure*}
\centering
\includegraphics[width=12cm,angle=0]{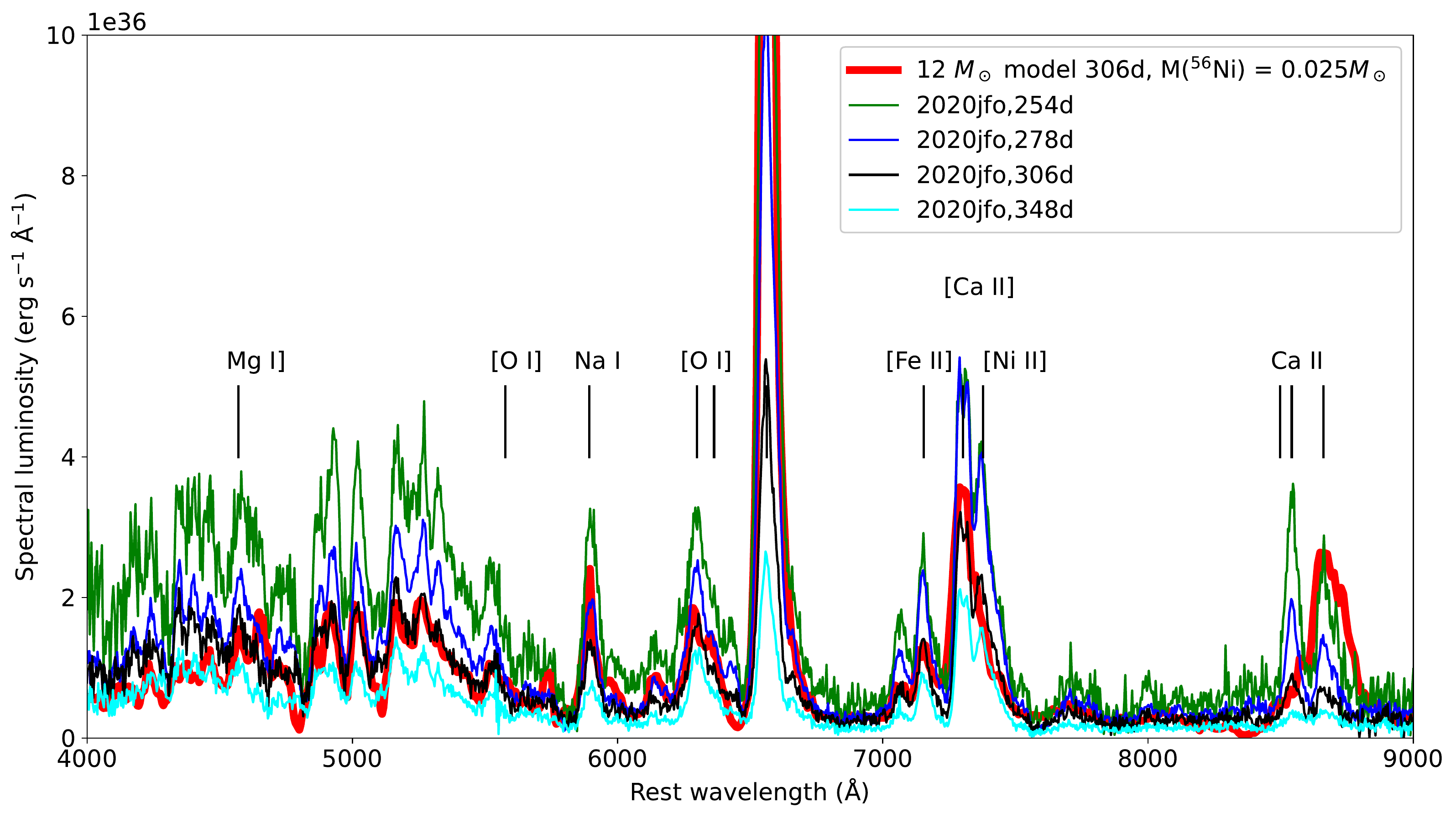} 
\caption{
\label{fig:2020jfonebularspectra}  
The four nebular spectra from NOT for SN 2020jfo are compared with
the nebular model for a 12 \Msun~explosion. The model is scaled by the ratio of $^{56}$Ni masses in SN 2020jfo (0.025 \Msun) and the model (0.062 \Msun), as well as a factor of 0.5 to account for a significantly lower degree of gamma-ray trapping in SN 2020jfo.
The most conspicuous emission lines are marked:
\ion{Mg}{I}]~$\lambda4571$, 
[\ion{O}{I}]~$\lambda5577$,  
[\ion{O}{I}]~$\lambda\lambda6300$, 6364, 
\ion{Na}{I D}, H$\alpha$, [\ion{Fe}{II}]~$\lambda7155$, [\ion{Ni}{II}]~$\lambda7378$, [\ion{Ca}{II}]~$\lambda\lambda$7291, 7323, and the \ion{Ca}{II} NIR triplet. 
}
\end{figure*}

In Fig.~\ref{fig:2020jfonebularspectra} we zoom in on the four high-quality 
late-time spectra of SN 2020jfo taken with the NOT.  These were 
obtained over a period when the SN was 250--350 days old, and the spectroscopic 
evolution over that time range is very slow. 
Overall, it is a textbook example of a normal SN~II spectrum
\citep{2017hsn..book..795J}, dominated by Balmer lines that still show P~Cygni absorption components, but also strong emission lines of calcium and oxygen.  
The spectra are indeed similar to those seen in many other nebular
CC~SNe, like the famous SN 1987A or the well-studied SN 2012aw.

For SN 2020jfo, where we have no evidence of CSM interaction, we can compare the nebular emission-line luminosities with modeling to estimate the oxygen mass and thus the zero-age main sequence (ZAMS) mass of the star that exploded. We use the models and methodology advanced by \citet{Jerkstrand2012}, \citet{Jerkstrand2015A&A...573A..12J..latetimespectra}, and \citet{Jerkstrand2018}. The spectra are calibrated on an absolute scale using the photometry and corrected for extinction. Figure~\ref{fig:2020jfonebularspectra} shows the 12 $M_\odot$ model from the work of \citet{Jerkstrand2015ApJ...807..110JNi/Feconstraints}. In these comparisons, we have rescaled the model flux with the ratio of $^{56}$Ni mass inferred for SN 2020jfo (0.025 \Msun; see  Sects.~\ref{sect:2020jfolc} and \ref{sect:MCMC}) to the model $^{56}$Ni mass (0.062 M$_\odot$). We have in addition multiplied the model by a factor of 0.5 to account for the significantly earlier gamma-ray escape occurring in SN 2020jfo.

The 12 \Msun model makes the best fit to the key emission lines that diagnose the progenitor mass (oxygen, sodium, and magnesium), and suggests that the best ZAMS-mass match for SN 2020jfo is in the range 10--15 \Msun. H$\alpha$ is too strong in the model, which is consistent with an unusually low hydrogen envelope mass in this SN, likely also causing the short plateau. Having a relatively low-mass He core progenitor that has lost a large part of its H envelope would be most naturally consistent with binary mass loss.

However, it is also noteworthy that the metal emission lines do not appear broader than in the model. With a low envelope mass, one would expect a low He core mass and a quite high explosion energy, as inferred in Sect.~\ref{sect:MCMC}, so the metals should have an unusually high expansion velocity. Thus, no fully self-consistent scenario is established.

Finally, note that the Ca~II NIR triplet mismatch is a well-known shortcoming of the models. The models give scattering of \ion{Ca}{II} $\lambda\lambda$8498, 8542 into the \ion{Ca}{II} $\lambda$8662 line --- but in most SNe~II this is not observed to happen.

\subsection{Stable nickel in SN 2020jfo}\label{sect:nebularnickellines}

\begin{figure*}
\centering
\includegraphics[width=12cm,angle=0]{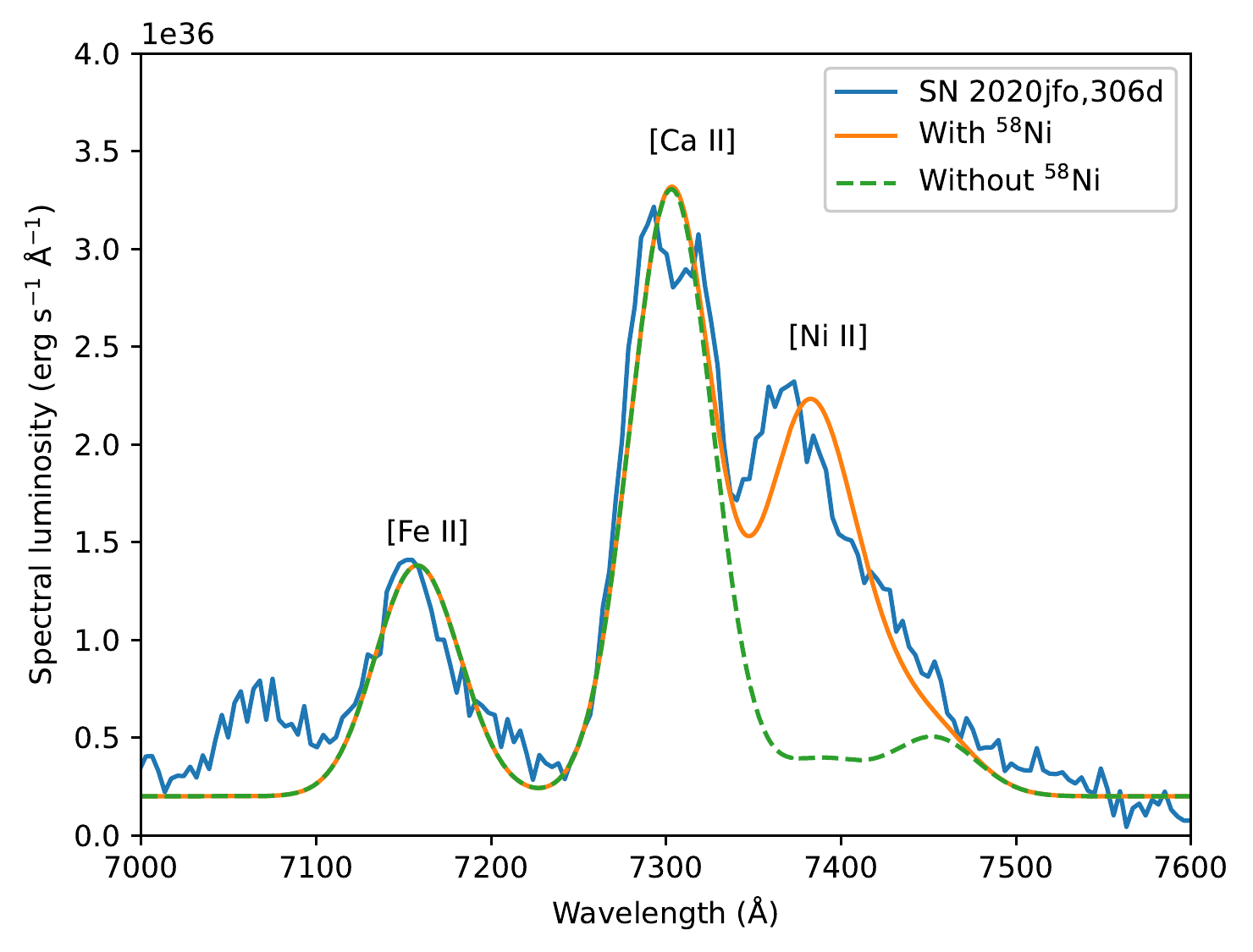} 
\caption{\label{fig:nickelline}  
The 7000--7600~\AA\ spectral range for SN 2020jfo at +306 days (blue lines). It has been demonstrated that this region is dominated by emission from [\ion{Fe}{II}], [\ion{Ni}{II}], and [\ion{Ca}{II}] \citep{Jerkstrand2015ApJ...807..110JNi/Feconstraints}. Shown in orange is the best-fit multiple-Gaussian model to the emission lines following the method of \citet{Jerkstrand2015ApJ...807..110JNi/Feconstraints}. This fit gives $L_{\rm Ni~II~ 7378}/L_{\rm Fe~II~7155}=1.7$, which maps to a mass ratio $M(\mbox{Ni})/M(\mbox{Fe})\approx 2$ times solar. The green dashed line shows the fit without contributions from stable nickel.}
\end{figure*}

The nebular spectra also show clear evidence for stable nickel in the form of the 
[\ion{Ni}{II}] $\lambda$7378 line (Fig.~\ref{fig:nickelline}). This line is not always seen in SNe II; a clearly observed line typically requires an unusually weak [\ion{Ca}{II}] doublet near 7300~\AA\ combined with an intrinsically strong nickel line. Here, it is plausible that the small hydrogen zone damps the calcium doublet, as much of that emission comes from this zone \citep{LiMcCray1993}. 

Stable nickel is an important diagnostic of the 
explosion mechanism. It is mainly composed of $^{58}$Ni, with a higher production most naturally being explained by burning and ejection of deeper-lying, more neutron-rich layers in the progenitor star \citep{Jerkstrand2015ApJ...807..110JNi/Feconstraints}. The iron comes from decayed $^{56}$Ni, which with its zero neutron excess has no such dependency.

{  We measure the relative luminosities for the lines following the approach of \citet[][their section 3.1.1]{Jerkstrand2015MNRAS.448.2482JMi/Fein2012ec}, with a simultaneous fit to several lines predicted from the models. Even if the exact line shape is not perfectly matched, we are confident about the line identifications. 
} The measured luminosity ratio at +306 days is $L_{\rm Ni II~7378}/L_{\rm Fe II~7155}=1.7$. The link between luminosity ratio and mass ratio depends on temperature, although quite weakly. The temperature can be estimated from the intrinsic line luminosity of [\ion{Fe}{II}] $\lambda7155$, the best-fitting value for $M({\rm Fe}) = 0.025$~\Msun\ being 2700~K.
A range of $T = 2500$--3000~K gives a mass ratio $M(\mbox{Ni})/M(\mbox{Fe}) = 1.7$--2.1, following the analysis method of \citet{Jerkstrand2015ApJ...807..110JNi/Feconstraints}.

\subsection{The nebular H$\alpha$ emission-line profiles}\label{sect:nebularlines}

If we focus on the strongest nebular emission line, H$\alpha$, we can see that it is relatively symmetric in SN 2020jfo, with FWHM $\approx 2510$~km~s$^{-1}$.
The P~Cygni absorption has a maximum at 
4000~km~s$^{-1}$, and also the red side of the emission line reaches that velocity. The maximum velocity in the red absorption component is 6000~km~s$^{-1}$, and this is actually consistent with the red shoulder of the emission line. 
Since we were able to properly model the nebular spectrum of SN 2020jfo in Sect.~\ref{sect:nebularoxygenlines} without any input from CSM interaction, we will adopt the H$\alpha$ profile for this SN as a benchmark with which to compare our other two SNe, to highlight how CSM interaction (as evident in the LC evolution) can manifest itself in the line profiles.

\begin{figure*}
\centering
\includegraphics[width=12cm,angle=0]{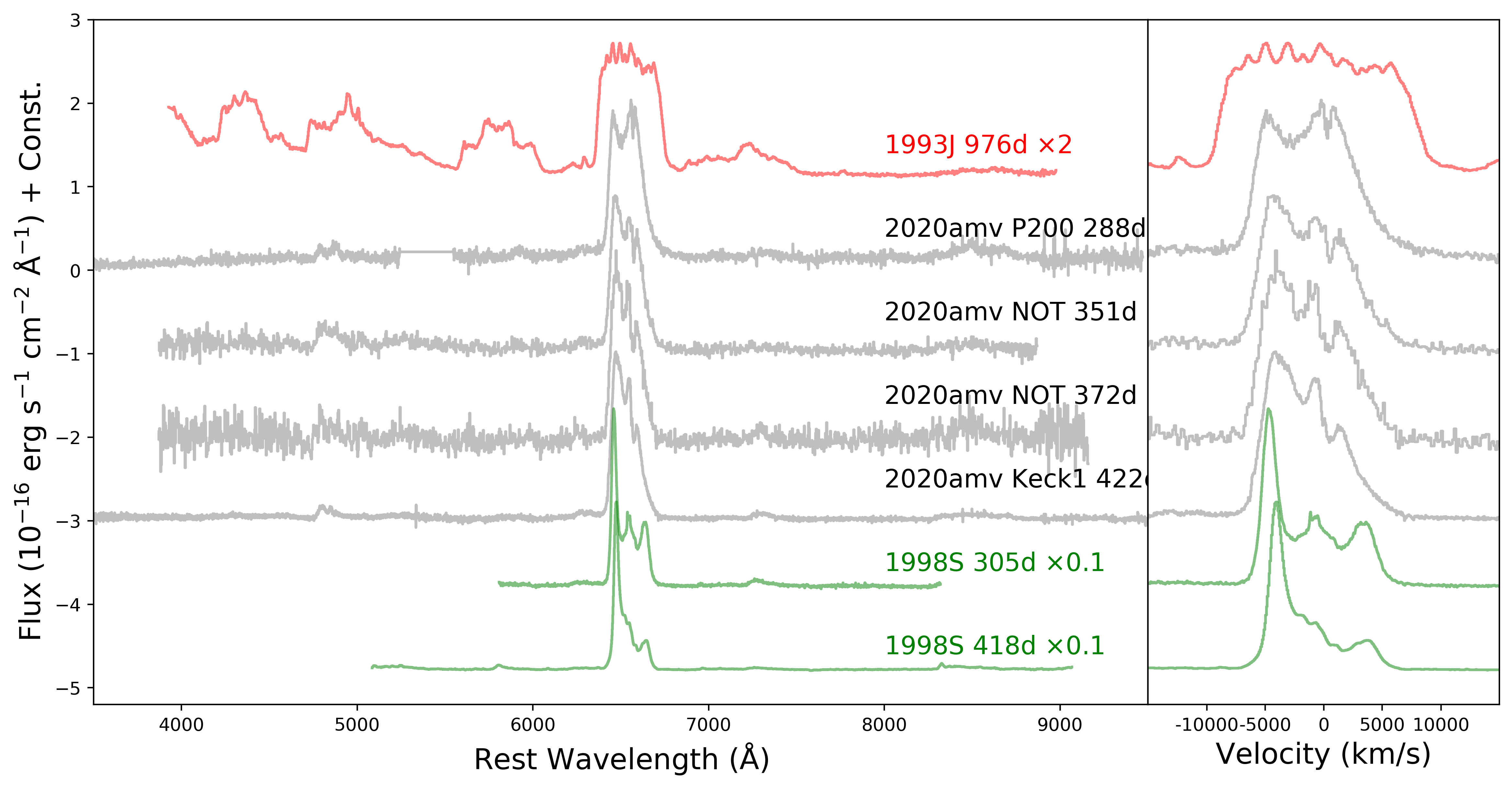} 
\caption{\label{fig:nebular2020amv}  
Nebular spectra of SN 2020amv. Left: The full range of the optical spectra, compared with SN 1993J (top, at 976 days, from \citealt{Matheson1993J}) which also showed a boxy line profile, and with a sequence of spectra of SN 1998S (305 to 418 days, from \citealt{Pozzo1998S}). Right: Zoom-in on the H$\alpha$ profile in velocity space. The profile is overall flat-topped and box-shaped in the first epoch, indicating emission from a thin dense shell formed when ejecta interact with the CSM. The later spectra  show the development of an increasing overall asymmetry and structure, with a clear blueshifted horn at $-5000$~km~s$^{-1}$. The evolution, with the suppression of the red side, is reminiscent of the evolution seen in SN 1998S.}
\end{figure*}

SN 2020amv is also a SN~II, but the nebular spectrum is strikingly different. We acquired two SEDM spectra at an age of 240 days, motivated by the endurance of the SN LC.
The SEDM spectra revealed little more than the H$\alpha$ emission line, but it caught our attention since it was very broad. The resolution of SEDM is low, but the width of the emission line stood out; we estimated a FWZI (full width at zero intensity) of $\sim 17,000$~km~s$^{-1}$. 
This prompted four more nebular spectra with larger telescopes. The first of these, from P200, is spectacular. It shows mainly H$\alpha$, but also H$\beta$ with the same line profile as well as the Ca II NIR triplet. The H$\alpha$ line profile displays three distinct peaks, similar to the late-time spectra of SN 1998S \citep[][their fig. 5]{Pozzo1998S} and SN 1993J \citep[see e.g.,][their fig. 11]{Matheson1993J}.
In Fig.~\ref{fig:nebular2020amv} we show the four last nebular spectra of SN 2020amv, compared to those of other SNe. The H$\alpha$ line in the P200 spectrum has FWHM $\approx 9800$~km~s$^{-1}$ (not measured with a Gaussian fit; FWZI $\approx 14,000$~km~s$^{-1}$). For the similar SN 1993J, the interpretation for the shape of the line profile
was CSM interaction with emission originating in a dense thin shell, and this obviously also applies to SN 2020amv.

The P200 H$\alpha$ line profile is asymmetric. The structure on top of the rectangular, boxy, and flat-topped line profile predicted from a thin shell can be interpreted either as evidence for an overall asymmetric geometrical configuration such as a ring-like structure
\citep[as mentioned by][]{Pozzo1998S}, or alternatively seen as small-scale structure as due to clumps, as suggested for the forbidden oxygen lines in SN 1993J by \citet{Spryromilio}.
In the P200 spectrum (Fig.~\ref{fig:nebular2020amv}) 
the blue-horn emission is shifted by
$5000$~km~s$^{-1}$ with respect to the galaxy rest frame. The same structure remains in the later NOT spectra, but it can be seen that the relative strengths of the features is changing. The final Keck spectrum spectacularly shows the three horns at $\sim -4240$, $-730$, and +1400~km~s$^{-1}$. 
The evolution of the three components is particularly conspicuous between the P200 spectrum and the final Keck spectrum.
We see that the emission line becomes more asymmetric with time, and that the blue horn is dominating the line profile at the last epoch, or rather that the red-most side of the line profile is suppressed. This is similar to the case of SN 1998S (Fig.~\ref{fig:nebular2020amv}).
The line-profile evolution of SNe 1993J and 1998S were discussed in some detail by \cite{Fransson2005ApJ...622..991F}. Several models for the geometry and dust distributions were tested, but none could convincingly explain the profiles of SN 1998S. The central component of the line profile would require an additional emission component ---
but the possibility that the central horn of the H$\alpha$ line in SN 1998S was affected by host contamination was also mentioned \citep{Fransson2005ApJ...622..991F}. In this regard, we note that the central component in the SN 2020amv Keck spectrum is real and clearly resolved at 
900~km~s$^{-1}$.
The host-galaxy H$\alpha$ line from the same epoch has FWHM $\approx 240$~km~s$^{-1}$. 
The offset between the two lines is 730~km~s$^{-1}$, with the central SN component blueshifted.
Overall, the remarkable resemblance and the similar line-profile evolution between SN 1998S and SN 2020amv argue for a generic scenario rather than a fine-tuned geometry and dust distribution.

Looking finally also at SN 2020jfv, the nebular emission-line spectrum
is somewhat of an intermediate case between the two above-mentioned
SNe. This SN also caught our attention given the photometric behaviour
(rebrightening).
Late-time SEDM spectra showed an unusual behaviour --- a stripped-envelope SN transforming into a SN~II --- which made us activate larger telescopes. The NOT data revealed a Balmer-dominated spectrum, but also clear [\ion{O}{I}] $\lambda\lambda$6300, 6364 emission and what is likely [\ion{Ca}{II}] at 7300~\AA. The later Keck spectrum confirms this, and is even more dominated by H$\alpha$ and H$\beta$; we measure a flux ratio of H$\alpha$ to [\ion{O}{I}] $\lambda\lambda$6300, 6364 of $\sim6.5$. The Ca~II NIR triplet is also very weak. The peak of the H$\alpha$ line profile is unfortunately damaged by a cosmic-ray hit in the high-S/N Keck spectrum. {  The final NOT spectrum, obtained after solar conjunction, shows basically only the Balmer lines.}

\begin{figure*}
\centering
\includegraphics[width=12cm,angle=0]{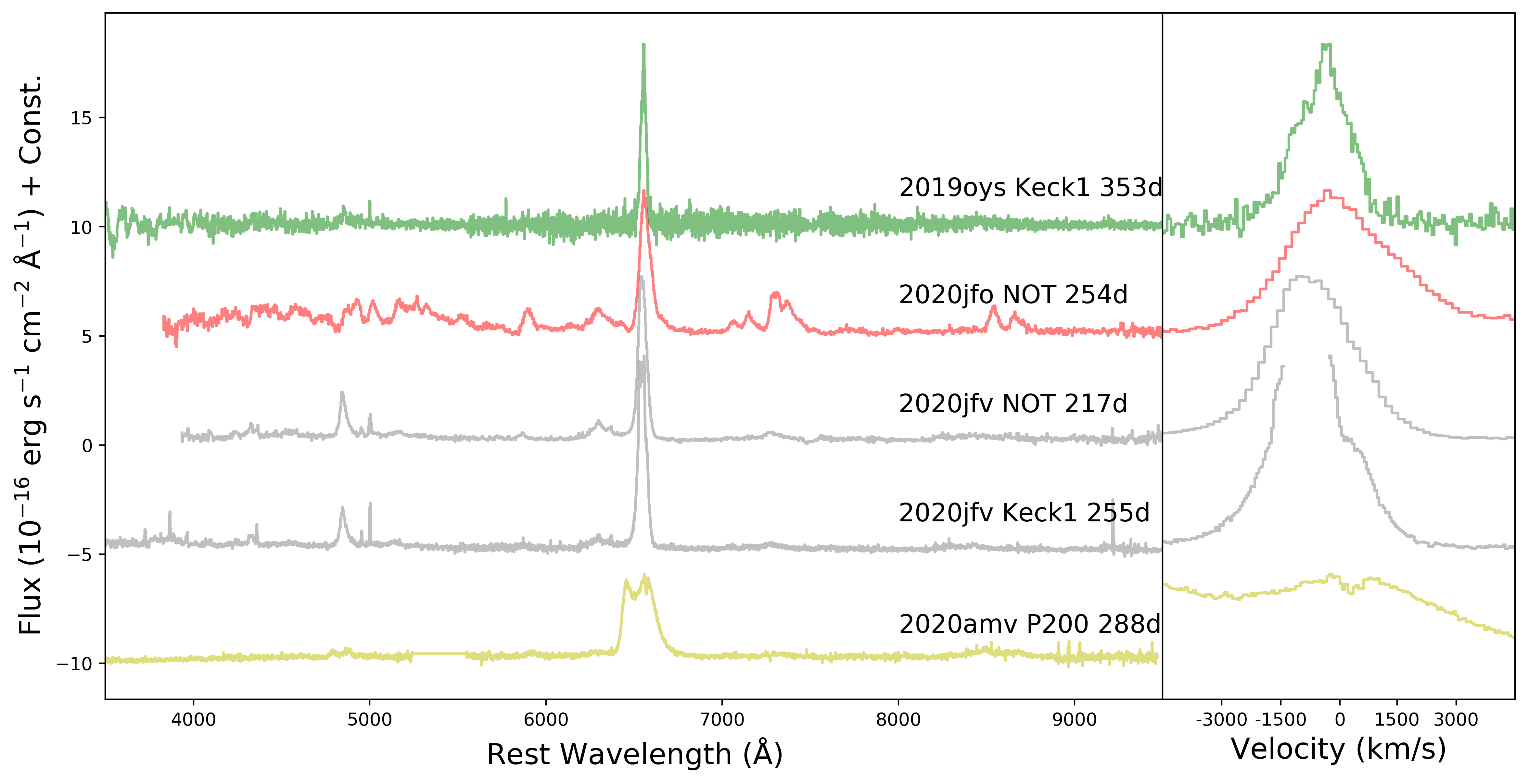}  
\caption{\label{fig:nebularspec2020jfv}  
Nebular spectra of SN 2020jfv compared with the normal SN 2020jfo, the CSM interacting SN 2020amv, and the transitional SN 2019oys \citep{sollerman2020}. The right panel shows a zoom-in on the H$\alpha$ line profile in velocity space. 
H$\alpha$ dominates all these spectra, but the spectral evidence for CSM interaction is more subtle in SN 2020jfv, with a small structure visible at the red end of the line in the high-S/N Keck spectrum. The peak of this line is affected by a cosmic-ray hit and is therefore not shown.
}
\end{figure*}

The line profile of H$\alpha$ in SN 2020jfv has FWHM $\approx 2500$~km~s$^{-1}$, which is virtually the same as for H$\beta$ (2800~km~s$^{-1}$, corrected for instrumental resolution; the host-galaxy lines are 750~km~s$^{-1}$).
Whereas the host lines appear at $z=0.017$, both H$\alpha$ and H$\beta$ are blueshifted by 
800 km~s$^{-1}$ from this. The extinction-corrected Balmer ratio is $\sim5$, indicating shock interaction.
In Fig.~\ref{fig:nebularspec2020jfv} we compare two late spectra of SN 2020jfv with our best spectrum of SN 2020jfo, the P200 spectrum of SN 2020amv, and also with SN 2019oys which was a stripped-envelope SN that transformed into a SN~IIn owing to late CSM interaction \citep{sollerman2020}. We can see that the spectrum of SN 2020jfv is different from all of these comparison objects. The right-hand panel of the figure zooms in on H$\alpha$ in velocity space, and we note that the intermediate-width emission line of SN 2020jfv actually has structure on the red side of the line profile. This is significantly more subtle evidence of CSM interaction than for SN 2020amv. 

\subsection{Uncertainties}\label{sec:uncertainties}
{  

\subsubsection{Distance and extinction}
For the analysis and discussion so far we made use of the provided distances and assumed no additional extinction from the host galaxies of our SNe. In this section we discuss what the main uncertainties are provided the errors in the distance estimates, and discuss how the derived parameters would change if more extinction in the host galaxies would dim and redden the light from the SNe. These are the observational main caveats for most SN studies.

In Sect.~\ref{sec:obs} we declared that the distance to M61 is uncertain and adopted
a distance modulus of $30.81\pm0.20$ mag. This estimate comes from the Expanding Photosphere Method for a Type II SN \citep{bosekumar}, and is also consistent with the peculiar motion corrected luminosity distance derived from standard cosmology and the observed redshift from NED. Such an uncertainty directly translates to a 20\% error in the nickel mass estimates. There are, however, other estimates of the distance to M61 that are even larger \citep[by almost 30\%, ][]{2015ApJ...799..215P}. 

Extinction is sometimes even more difficult to determine. We have assumed no host galaxy 
extinction for the three SNe, mainly based on their blue colors.
For SN 2020jfo there is some
evidence for narrow \ion{Na}{I~D} lines in the Lick spectrum, where we can estimate an equivalent width of $\lesssim0.7$~\AA~for the doublet. 
For example using \cite{Taubenberger2006} with A$_{V} = 3.1 \times 0.16~\times \rm{EW(NaID)}$ would give 0.3 mag of extinction in the optical.  In most regards such a 30\% increase in flux is similar to adopting a larger distance, as discussed above. 

With more color information available from the pre-explosion {\it HST} imaging, such a reddening would also be of importance for the progenitor conclusions. However, as noted in Sect.~\ref{sec:preexplosionimaging} we detect the probable progenitor only in the reddest band, and it is fainter than expected by more than a magnitude. 
A host extinction correction of the magnitude suggested above would not resolve this issue.

A main effect from the uncertainty in distance and reddening is on the actual mass of radioactive nickel as mentioned above, the effects of both a larger distance and some host extinction could potentially increase the estimate from 0.025 \Msun to 0.04 \Msun. Other parts of the analysis is less sensitive. For example the modeling of the nebular spectra relies on relative line luminosities for the $^{58}$Ni analysis, which is virtually independent of both distance and extinction. Also the ZAMS mass estimate from the oxygen mass is derived by scaling the oxygen luminosity with the derived nickel mass, and both are similarly affected by the uncertainties. 

Shortly, also for SNe 2020bmv and 2020jfo the discussion is similar as above, but there is not much analysis that is severely affected. The distance estimates are deduced from the redshifts, and at 74 and 200 Mpc the effects of peculiar velocities is smaller (V$_{\rm{pec}} = 150~$km~s$^{-1}$ gives at most 6\% uncertainty in flux), and is instead dominated by the everpresent uncertainty in the Hubble constant ($\pm3$~km s$^{-1}$ Mpc$^{-1}$ gives 9\% in flux). 
For these two peculiar SNe, the fact that the colors are not red is not much evidence against host galaxy reddening. None of our spectra show evidence for narrow ISM lines (SN 2020jfv have no constraining observations, whereas the NTT spectrum of SN 2020amv would reveal a Na ID of the same strengths as potentially present for SN 2020jfo).
However, our discussion for these two SNe mostly concerned the line profiles and their evolution, and is not affected by these uncertainties.

\subsubsection{Methodology}

We can also briefly discuss the different methodologies used to infer the properties of SN\,2020jfo and its progenitor. We have been able to use relatively well-established procedures to infer properties of the progenitor star from complementary routes such as pre-explosion progenitor imaging, bolometric light-curve modeling and nebular NLTE emission-line analysis. Whereas the derived oxygen mass implies a progenitor of ZAMS mass around 12~\Msun, the short LC plateau indicates a lower ejecta mass and the progenitor detection is fainter still. This could imply interesting properties of the supernova, such as circumstellar dust around the progenitor destroyed in the explosion and extensive mass-loss affecting the hydrogen envelope, as mentioned in the next section. The tension could also indicate that some of the many assumptions inherent in these methods may not hold. More well-studied SNe where several lines of investigation can be applied are needed.
}

\section{Summary, interpretation and conclusions}\label{sec:conclusions}

In this paper we have presented the discovery, classification, and follow-up observations of the Type II SN 2020jfo, which exploded in the nearby spiral galaxy M61 in May 2020. We presented optical, NIR, and NUV photometry, as well as spectroscopy, for the first year of this transient, which was also followed by many other astronomers (both professional and amateur) around the globe. Even though the site was covered by PTF and ZTF for 11~yr prior to explosion, we did not detect any pre-SN outbursts down to a magnitude of $-11$. 

The supernova has a well-constrained explosion epoch and rose to a maximum of $M^{\rm{peak}}_r = -16.5$ mag in $\sim5$ days. It was not detected at any significance in X-rays using \swift, but the UV LC was well covered. The plateau length of 65 days is on the short side of the SN~IIP distribution. Using simple modeling we estimate an ejecta mass of $M_{\rm ej} \approx 5$ \Msun, while the mass of radioactive nickel estimated from the late-time tail of the LC is $0.025$ \Msun, {  although uncertainties in distance and host extinction could increase this by $\sim$50\%}. The spectroscopic sequence of SN 2020jfo was largely obtained with robotic low-resolution spectrographs, but reveals normal SN~II evolution dominated by Balmer lines with P~Cygni profiles. This sort of sequence can routinely be achieved with SEDM-like spectrographs. 

We also secured two epochs of spectropolarimetry. The S/N is rather low, but there is evidence of both continuum and line polarization. The interpretation is uncertain, but 
the modest 
continuum polarization may be due to an asymmetric distribution of radioactive elements, or of an asymmetric electron density structure. The higher line polarization suggests that Ca is not uniformly distributed within the ejecta and likely exists in clumps. 

For the later nebular phases, we rely on ToO triggers on larger telescopes. The sequence of spectra of SN 2020jfo from the NOT displays slow evolution in the nebular phase. Line-flux measurements of the calibrated spectra indicate, when compared to detailed NLTE models, that the exploding progenitor had a ZAMS mass of $\sim12$ \Msun. 
Nebular emission line analysis also revealed a high abundance of stable nickel ($^{58}$Ni), with 
a mass ratio $M(\mbox{Ni})/M(\mbox{Fe}) \approx 2$.
Only a handful of CC~SNe have previous estimates of this mass ratio \citep{Maeda2007,Jerkstrand2015MNRAS.448.2482JMi/Fein2012ec, Terreran2016,Tomasella2018,MullerBravo2020}. The picture so far points to a value of around solar being most common --- this corresponds to burning and ejection of oxygen-rich layers. SN 2020jfo joins a small group of SNe having an enhanced ratio of 2--4 times solar; others in this group are the normal Type IIP SN 2012ec \citep{Jerkstrand2015MNRAS.448.2482JMi/Fein2012ec} and the broad-lined Type Ic SN 2006aj \citep{Maeda2007}. 
The result for SN 2020jfo adds another important piece to the puzzle. With its moderate progenitor mass, there appears currently to be no simple dependency on progenitors mass, but rather that both low-mass and high-mass stars can achieve ejection of neutron-rich material.

Pre-explosion {\sl HST\/} imaging reveal a putative 
source at the site of SN 2020jfo.
{  The exact location was obtained with ground-based adaptive optics imaging, and confirmed with post-explosion {\it HST} imaging.}
The star is 
only detected in the reddest band with an absolute magnitude of
$M_{\rm F814W} \approx -5.8$, and nondetections in the bluer bands. This is fainter than expected and might imply circumstellar dust that was later destroyed by the SN explosion.
The notably short plateau of SN 2020jfo implies that partial stripping of the envelope occurred at some episode before explosion, either through a stellar wind or mass exchange with a binary companion. Such mass loss would result in CSM, which could be consistent with the possible presence of circumstellar dust. 
If the CSM were set up by a vigorous pre-SN wind,
it was not accompanied by any luminous outburst (Sect. \ref{sec:preexplosionimaging}).

We have examined BPASS binary models \citep{Eldridge2019} for the mass range of 10 -- 15 \Msun, as indicated for the ZAMS mass by the nebular line analysis. Although a number of models terminate near the observed $M_{\rm F814W}$ (without the presence of further circumstellar extinction),  the very low amount of H mass in these models at the time of explosion 
is more consistent with what we would expect for a SN~IIb than a SN~IIP progenitor. The reason for the faint, red possible detection of the SN 2020jfo progenitor, given the properties of the SN itself, 
therefore remains unknown and requires further study. 
Ultimately, the identification of the progenitor candidate must be confirmed by revisiting the site when the SN has sufficiently faded.

We have compared SN 2020jfo with two other ZTF SNe, all three being dominated by Balmer lines in the nebular spectra. SN 2020amv is also a SN~II, but with a different LC and spectral evolution. The long-lived bumpy LC is a telltale sign of interaction with CSM. This is confirmed both by strong flash-spectroscopy features at early times, and a broad, square-shaped nebular emission-line profile in H$\alpha$, which furthermore consists of several components. The interpretation for such a line profile is emission from a cold dense shell developed by the reverse shock from interaction with dense CSM that the ejecta run into at later phases. CSM interaction is likely also the reason why SN 2020jfv rebrightens hundreds of days past explosion. This SN~IIb initially did not have strong signatures of hydrogen, whereas the nebular spectrum is completely dominated by Balmer lines. The line shapes are somewhat intermediate between those discussed for the other two SNe above. Instead of evidence for CSM interaction in terms of a broad, flat-topped line profile, we see an intermediate-width asymmetric line, where optical-depth effects likely play a role in shaping the line. Interest in both of the latter targets was raised only after we realised their slowly evolving and rebrightening light curves, and in both cases initial SEDM data convinced us to also obtain spectra at larger telescopes. SEDM can thus operate both as a classification machine and as a science data provider (as for SN 2020jfo), but also as a way to investigate whether additional observations should be undertaken for any given transient. In SN 2020jfo itself, we see no evidence for CSM interaction, neither in the LC nor in the spectral evolution. However, based on the LC fit to the short plateau, and to the fast-declining late-time bolometric LC, we find evidence for a relatively low mass of hydrogen ejecta. Given the ZAMS mass derived from the nebular-line analysis, this points to a significant amount of mass loss. Since this CSM is not affecting the observed properties of the SN, we suspect that this period of mass loss must have occurred at a substantially earlier phase of the progenitor's evolution.

Large surveys such as ZTF are now routinely discovering thousands of SNe. Part of the observations that were previously cumbersome and expensive to obtain, such as continuous LCs in several bands from early to late times and decent numbers of low-resolution spectra, are now virtually automatically available to the astronomical community --- in particular, for nearby events such as SN 2020jfo. The challenge has moved to being able to digest and publish the available data, and to detect, select, and monitor events of particular interest. SNe 2020amv and 2020jfv were both unusually long lived with LCs that rebrightened. Dedicated spectroscopic follow-up observations were required to confirm CSM interaction as the powering mechanism for these objects.

\begin{acknowledgements}
    We thank the staffs of the various observatories where data were obtained for their excellent assistance.
    Based on observations obtained with the Samuel Oschin Telescope 48-inch and the 60-inch Telescope at the Palomar Observatory as part of the Zwicky Transient Facility project. ZTF is supported by the U.S. National Science Foundation (NSF) under grant AST-1440341 and a collaboration including Caltech, IPAC, the Weizmann Institute for Science, the Oskar Klein Center at Stockholm University, the University of Maryland, the University of Washington, Deutsches Elektronen-Synchrotron and Humboldt University, Los Alamos National Laboratories, the TANGO Consortium of Taiwan, the University of Wisconsin at Milwaukee, and Lawrence Berkeley National Laboratories. 
    ZTF-II is supported by the National Science Foundation under Grant No. AST-2034437 and a collaboration including Caltech, IPAC, the Weizmann Institute for Science, the Oskar Klein Center at Stockholm University, the University of Maryland, Deutsches Elektronen-Synchrotron and Humboldt University, the TANGO Consortium of Taiwan, the University of Wisconsin at Milwaukee, Trinity College Dublin, Lawrence Livermore National Laboratories, and IN2P3, France. Operations are conducted by COO, IPAC, and UW. 
The SED Machine is based upon work supported by the NSF under grant  AST-1106171. The ZTF forced-photometry service was funded under the Heising-Simons Foundation grant \#12540303 (PI M. Graham). 
This work was supported by the GROWTH project 
funded by the NSF under PIRE grant 1545949. 
The Oskar Klein Centre was funded by the Swedish Research Council.
Gravitational Radiation and Electromagnetic Astrophysical Transients (GREAT) is funded by the Swedish
Research council (VR) under Dnr 2016-06012. ECK is further supported by
the Wenner-Gren Foundations.
Much support to OKC involvement in ZTF was provided by the Knut and Alice Wallenberg foundation.
Partially based on observations made with the Nordic Optical Telescope, operated by the Nordic Optical Telescope Scientific Association at the Observatorio del Roque de los Muchachos, La Palma, Spain, of the Instituto de Astrofisica de Canarias. Some of the data presented here were obtained with ALFOSC. 
Based on observations made with the NASA/ESA {\it Hubble Space Telescope}, obtained from the data archive at the Space Telescope Science Institute (STScI). STScI is operated by the Association of Universities for Research in Astronomy, Inc. under NASA contract NAS 5-26555. {  Support for program GO-16179 was provided by NASA through a grant from STScI. We thank T. de Jaeger, O. D. Fox, P. Kelly, N. Smith, S. Vasylyev, and W. Zheng for their assistance with this program.}
Some of the data presented herein were obtained at the W. M. Keck
Observatory, which is operated as a scientific partnership among the
California Institute of Technology, the University of California, and
NASA; the observatory was made possible by the generous financial
support of the W. M. Keck Foundation. The authors wish to recognize and acknowledge the very significant cultural role and reverence that the summit of Maunakea has always had within the indigenous Hawaiian community.  We are most fortunate to have the opportunity to conduct observations from this mountain.
A major upgrade of the Kast spectrograph on the Shane 3~m telescope at Lick Observatory was made possible through generous gifts from the Heising-Simons Foundation as well    
as William and Marina Kast. Research at Lick Observatory is partially supported by a generous gift from Google.
Palomar Gattini-IR (PGIR) is generously funded by Caltech, the Australian National University, the Mt. Cuba Foundation, the Heising-Simons Foundation, and the Binational Science Foundation. PGIR is a collaborative project among Caltech, the Australian National University, University of New South Wales, Columbia University, and the Weizmann Institute of Science.
Y.-L.K. has received funding from the European Research Council (ERC) under the European Unions Horizon 2020 research and innovation program (grant 759194 -- USNAC).
 M.C. acknowledges support from the NSF through grant PHY-2010970.
 Funding for A.V.F.'s research group at U.C. Berkeley was provided by the Christopher R. Redlich Fund and the Miller Institute for Basic Research in Science (where A.V.F. is a Senior Miller Fellow). This work was partially funded by {\it Kepler/K2} grant J1944/80NSSC19K0112, {\it HST} grant GO-15889 from STScI, and STFC grants ST/T000198/1 and ST/S006109/1. A.J. acknowledges funding from the European Research Council (ERC) through Starting Grant 803189.
  This work has made use of data from the Asteroid Terrestrial-impact Last Alert System (ATLAS) project. The Asteroid Terrestrial-impact Last Alert System (ATLAS) project is primarily funded to search for near-Earth asteroids through NASA grants NN12AR55G, 80NSSC18K0284, and 80NSSC18K1575; byproducts of the asteroid search include images and catalogues from the survey area. The ATLAS science products have been made possible through the contributions of the University of Hawaii Institute for Astronomy, the Queen’s University Belfast, STScI, the South African Astronomical Observatory, and The Millennium Institute of Astrophysics (MAS), Chile.

\end{acknowledgements}

\clearpage

\bibliography{m61}

\clearpage

\begin{table*}
\caption{Supernova and host-galaxy properties}              % title of Table
\label{table:SNproperties}       % is used to refer this table in the text
\centering                                      % used for centering table
\resizebox{\textwidth}{10mm}{ % resize table, width and height
\begin{tabular}{l c ccccc c}          % centered columns (4 columns)
\hline\hline                        % inserts double horizontal lines
IAU name&  ZTF name & Explosion date  & SN Type & Redshift & Host & Distance & $E(B-V)_{\rm MW}$ \\    
 &   & (JD)  &  &  &  & (Mpc) & (mag) \\  
% table heading
\hline                                   % inserts single horizontal line
SN 2020amv & ZTF20aahbamv & 2458871.22 & IIn & 0.0452  & WISEA J084940.38+301115.5 & 200.3 & 0.032 \\
SN 2020jfo & ZTF20aaynrrh & 2458975.20 & II & 0.00522 & M61 & 14.5 & 0.020 \\
SN 2020jfv & ZTF20abgbuly & 2458969.51 & IIb$\rightarrow$IIn & 0.017 & WISEA J230635.97+003641.9 & 73.8 & 0.045 \\      
\hline                                             
\end{tabular}}
\end{table*}

\begin{table*}
\caption{Summary of ZTF $gri$-band and ATLAS $co$-band observations (forced photometry) of SN 2020jfo.
ATLAS and ZTF data are binned in 3 and 0.5 days (respectively).
%using \href{https://gist.github.com/thespacedoctor/86777fa5a9567b7939e8d84fd8cf6a76}{atlas-fp}. 
The full table will be made available in machine-readable format.
} 
\label{tab:2020jfo_phot} 
\centering 
\begin{tabular}{lcc} 
\hline\hline 
Observation &  filter & mag \\  
Date (JD) &   &  \\  \hline          
2458975.70 & r & $16.00 \pm 0.01$ \\
2458975.75 & g & $15.40 \pm 0.01$ \\
2458976.37 & i & $15.49 \pm 0.03$ \\
2458976.94 & o & $15.06 \pm 0.00$ \\
2458977.41 & i & $15.06 \pm 0.00$ \\
2458978.42 & i & $14.84 \pm 0.00$ \\
2458978.69 & g & $14.47 \pm 0.00$ \\
2458978.75 & r & $14.52 \pm 0.00$ \\
2458978.96 & o & $14.64 \pm 0.00$ \\
2458980.41 & i & $14.66 \pm 0.00$ \\
2458980.86 & o & $14.52 \pm 0.00$ \\
2458981.70 & g & $14.45 \pm 0.00$ \\
2458981.75 & r & $14.46 \pm 0.00$ \\
2458982.87 & c & $14.51 \pm 0.00$ \\
2458983.39 & i & $14.69 \pm 0.00$ \\
2458984.90 & o & $14.58 \pm 0.00$ \\
2458986.69 & g & $14.56 \pm 0.00$ \\
2458986.76 & r & $14.46 \pm 0.00$ \\
2458986.87 & c & $14.63 \pm 0.00$ \\
2458988.85 & o & $14.59 \pm 0.00$ \\
2458989.38 & i & $14.75 \pm 0.00$ \\
2458990.83 & c & $14.66 \pm 0.00$ \\
2458992.69 & g & $14.70 \pm 0.00$ \\
2458992.73 & r & $14.46 \pm 0.00$ \\
2458992.85 & o & $14.58 \pm 0.00$ \\
2458994.85 & c & $14.76 \pm 0.00$ \\
2458995.70 & g & $14.79 \pm 0.00$ \\
2458995.73 & r & $14.51 \pm 0.00$ \\
2458996.85 & o & $14.61 \pm 0.00$ \\
2458998.69 & g & $14.88 \pm 0.00$ \\
2458998.73 & r & $14.47 \pm 0.00$ \\
2458998.81 & o & $14.60 \pm 0.00$ \\
2459005.69 & g & $15.07 \pm 0.00$ \\
2459005.83 & o & $14.65 \pm 0.00$ \\
2459008.84 & o & $14.66 \pm 0.00$ \\
2459009.73 & g & $15.25 \pm 0.00$ \\
2459010.82 & o & $14.65 \pm 0.00$ \\
2459012.73 & i & $14.66 \pm 0.03$ \\
2459014.70 & g & $15.25 \pm 0.00$ \\
2459016.80 & o & $14.70 \pm 0.00$ \\
2459017.74 & r & $14.65 \pm 0.00$ \\
2459017.81 & i & $14.73 \pm 0.04$ \\
2459020.70 & g & $15.33 \pm 0.00$ \\
2459020.74 & r & $14.68 \pm 0.00$ \\
2459020.79 & o & $14.75 \pm 0.00$ \\
2459022.78 & c & $15.17 \pm 0.01$ \\
2459023.71 & g & $15.41 \pm 0.00$ \\
2459024.77 & o & $14.81 \pm 0.00$ \\
2459026.77 & o & $14.80 \pm 0.00$ \\
2459030.76 & o & $14.87 \pm 0.00$ \\
2459032.80 & o & $15.01 \pm 0.00$ \\
2459033.68 & r & $15.01 \pm 0.00$ \\
2459033.70 & g & $15.90 \pm 0.01$ \\
2459035.64 & o & $15.12 \pm 0.01$ \\
2459037.71 & g & $16.40 \pm 0.02$ \\
2459040.79 & o & $15.72 \pm 0.01$ \\
2459042.79 & c & $16.75 \pm 0.03$ \\
2459044.76 & o & $16.38 \pm 0.02$ \\
2459064.75 & o & $16.67 \pm 0.02$ \\
2459158.03 & r & $17.74 \pm 0.04$ \\
2459164.03 & r & $17.69 \pm 0.02$ \\
2459166.03 & r & $17.79 \pm 0.02$ \\
\end{tabular}
\hspace{1em}
\begin{tabular}{lcc} 
\hline\hline 
Observation &  filter & mag \\  
Date (JD) &   &  \\  \hline  
2459168.03 & r & $17.91 \pm 0.03$ \\
2459176.00 & r & $18.09 \pm 0.26$ \\
2459177.02 & g & $18.97 \pm 0.06$ \\
2459177.06 & r & $17.99 \pm 0.05$ \\
2459181.00 & g & $19.36 \pm 0.17$ \\
2459181.06 & r & $18.07 \pm 0.05$ \\
2459182.50 & o & $18.23 \pm 0.11$ \\
2459185.01 & r & $18.13 \pm 0.04$ \\
2459186.15 & o & $18.21 \pm 0.10$ \\
2459187.04 & g & $19.03 \pm 0.17$ \\
2459189.01 & r & $18.17 \pm 0.04$ \\
2459189.04 & g & $18.92 \pm 0.07$ \\
2459190.13 & o & $18.24 \pm 0.11$ \\
2459191.11 & o & $18.21 \pm 0.11$ \\
2459195.02 & r & $18.29 \pm 0.03$ \\
2459195.05 & g & $19.15 \pm 0.06$ \\
2459197.99 & c & $18.90 \pm 0.20$ \\
2459198.98 & r & $18.40 \pm 0.05$ \\
2459199.03 & g & $19.13 \pm 0.06$ \\
2459200.95 & g & $18.97 \pm 0.16$ \\
2459201.03 & r & $18.34 \pm 0.06$ \\
2459204.00 & r & $18.60 \pm 0.05$ \\
2459204.04 & g & $19.21 \pm 0.07$ \\
2459204.97 & g & $19.10 \pm 0.05$ \\
2459205.01 & r & $18.42 \pm 0.03$ \\
2459205.13 & c & $19.04 \pm 0.23$ \\
2459207.10 & c & $19.01 \pm 0.23$ \\
2459209.13 & c & $19.22 \pm 0.28$ \\
2459217.02 & r & $18.66 \pm 0.09$ \\
2459217.04 & g & $19.41 \pm 0.17$ \\
2459217.08 & o & $18.62 \pm 0.16$ \\
2459221.94 & r & $18.67 \pm 0.06$ \\
2459221.97 & g & $19.20 \pm 0.11$ \\
2459222.27 & o & $18.78 \pm 0.18$ \\
2459223.90 & g & $19.33 \pm 0.09$ \\
2459223.97 & r & $18.85 \pm 0.06$ \\
2459225.09 & c & $19.23 \pm 0.28$ \\
2459225.97 & g & $19.41 \pm 0.08$ \\
2459226.00 & r & $19.01 \pm 0.28$ \\
2459227.88 & r & $18.79 \pm 0.05$ \\
2459227.96 & g & $19.40 \pm 0.07$ \\
2459228.07 & c & $19.24 \pm 0.29$ \\
2459229.93 & r & $18.93 \pm 0.08$ \\
2459231.00 & c & $19.74 \pm 0.47$ \\
2459231.96 & g & $19.49 \pm 0.06$ \\
2459231.98 & r & $18.89 \pm 0.04$ \\
2459233.80 & c & $19.44 \pm 0.35$ \\
2459249.11 & o & $19.17 \pm 0.27$ \\
2459249.91 & r & $19.35 \pm 0.11$ \\
2459249.94 & g & $19.85 \pm 0.21$ \\
2459251.79 & i & $19.37 \pm 0.19$ \\
2459251.96 & r & $19.30 \pm 0.06$ \\
2459253.09 & o & $19.32 \pm 0.31$ \\
2459253.88 & g & $19.94 \pm 0.12$ \\
2459253.99 & r & $19.32 \pm 0.06$ \\
2459255.92 & r & $19.43 \pm 0.06$ \\
2459257.88 & g & $19.76 \pm 0.15$ \\
2459257.99 & c & $19.63 \pm 0.42$ \\
2459260.94 & c & $19.61 \pm 0.41$ \\
2459262.86 & r & $19.58 \pm 0.09$ \\
2459262.92 & g & $19.79 \pm 0.19$ \\
2459264.87 & r & $19.45 \pm 0.07$ \\
\end{tabular}
\hspace{1em}
\begin{tabular}{lcc} 
\hline\hline 
Observation &  filter & mag \\  
Date (JD) &   &  \\  \hline  

2459266.83 & i & $19.25 \pm 0.21$ \\
2459266.86 & r & $19.82 \pm 0.19$ \\
2459268.85 & r & $19.67 \pm 0.12$ \\
2459268.94 & g & $19.97 \pm 0.14$ \\
2459269.64 & o & $19.50 \pm 0.37$ \\
2459270.85 & r & $19.85 \pm 0.23$ \\
2459270.90 & g & $20.33 \pm 0.32$ \\
2459272.91 & g & $19.70 \pm 0.31$ \\
2459272.94 & r & $19.49 \pm 0.24$ \\
2459275.88 & r & $19.59 \pm 0.17$ \\
2459275.94 & g & $19.63 \pm 0.21$ \\
2459277.02 & o & $19.71 \pm 0.45$ \\
2459278.82 & g & $20.17 \pm 0.13$ \\
2459278.92 & r & $19.71 \pm 0.11$ \\
2459280.79 & r & $19.91 \pm 0.10$ \\
2459280.90 & g & $19.98 \pm 0.12$ \\
2459281.00 & o & $19.39 \pm 0.33$ \\
2459281.79 & r & $19.95 \pm 0.12$ \\
2459285.07 & o & $19.83 \pm 0.51$ \\
2459288.94 & o & $19.81 \pm 0.50$ \\
2459290.80 & g & $20.32 \pm 0.12$ \\
2459290.84 & r & $19.91 \pm 0.10$ \\
2459291.79 & g & $20.54 \pm 0.16$ \\
2459291.85 & r & $20.12 \pm 0.11$ \\
2459292.85 & r & $20.08 \pm 0.11$ \\
2459292.99 & o & $19.81 \pm 0.50$ \\
2459293.73 & r & $19.84 \pm 0.13$ \\
2459293.79 & g & $20.46 \pm 0.18$ \\
2459295.74 & r & $20.06 \pm 0.20$ \\
2459295.77 & g & $19.97 \pm 0.19$ \\
2459296.85 & r & $20.35 \pm 0.27$ \\
2459297.50 & o & $19.80 \pm 0.49$ \\
2459298.76 & g & $20.44 \pm 0.37$ \\
2459298.88 & r & $20.16 \pm 0.30$ \\
2459304.81 & r & $20.22 \pm 0.31$ \\
2459305.78 & g & $20.35 \pm 0.24$ \\
2459306.04 & o & $20.00 \pm 0.60$ \\
2459306.77 & r & $20.48 \pm 0.16$ \\
2459306.85 & g & $20.50 \pm 0.21$ \\
2459307.73 & r & $20.85 \pm 0.24$ \\
2459307.80 & g & $21.05 \pm 0.26$ \\
2459308.77 & g & $20.60 \pm 0.15$ \\
2459308.77 & r & $20.53 \pm 0.16$ \\
2459308.94 & c & $20.74 \pm 1.23$ \\
2459309.71 & g & $20.69 \pm 0.21$ \\
2459309.74 & r & $20.57 \pm 0.18$ \\
2459311.73 & g & $20.66 \pm 0.24$ \\
2459311.81 & r & $20.61 \pm 0.19$ \\
2459312.80 & r & $20.63 \pm 0.19$ \\
2459312.83 & g & $20.76 \pm 0.22$ \\
2459313.76 & g & $20.81 \pm 0.21$ \\
2459313.80 & r & $20.53 \pm 0.29$ \\
2459315.73 & r & $20.45 \pm 0.15$ \\
2459316.81 & r & $20.77 \pm 0.22$ \\
2459316.82 & g & $20.77 \pm 0.17$ \\
2459317.71 & r & $20.53 \pm 0.15$ \\
2459319.73 & r & $20.43 \pm 0.20$ \\
2459319.76 & g & $20.81 \pm 0.26$ \\
2459321.75 & g & $21.10 \pm 0.25$ \\
2459321.78 & r & $20.75 \pm 0.19$ \\
2459322.95 & o & $19.96 \pm 0.58$ \\
\end{tabular}
\end{table*}

\begin{table*}
\caption{Summary of ZTF $gri$-band and ATLAS $co$-band observations of SN 2020amv.
ATLAS and ZTF data are binned in 3 days. The full table will be made available in machine-readable format.
} 
\label{tab:2020amv_phot} 
\centering 
\begin{tabular}{lcc} 
\hline\hline 
Observation &  filter & mag \\  
Date (JD) &   &  \\  \hline          
2458871.72 & g & $18.71 \pm 0.04$ \\
2458871.77 & r & $19.02 \pm 0.04$ \\
2458871.89 & c & $18.76 \pm 0.18$ \\
2458873.98 & o & $18.25 \pm 0.11$ \\
2458874.77 & r & $17.99 \pm 0.02$ \\
2458874.85 & g & $17.63 \pm 0.01$ \\
2458875.99 & c & $17.64 \pm 0.06$ \\
2458877.78 & r & $17.72 \pm 0.02$ \\
2458877.94 & o & $17.74 \pm 0.07$ \\
2458878.76 & g & $17.47 \pm 0.02$ \\
2458880.82 & r & $17.59 \pm 0.01$ \\
2458882.96 & o & $17.65 \pm 0.06$ \\
2458883.90 & g & $17.46 \pm 0.02$ \\
2458884.70 & r & $17.60 \pm 0.03$ \\
2458890.70 & r & $17.83 \pm 0.05$ \\
2458890.76 & g & $17.80 \pm 0.04$ \\
2458893.76 & r & $17.77 \pm 0.02$ \\
2458894.70 & g & $17.82 \pm 0.02$ \\
2458898.74 & g & $18.04 \pm 0.02$ \\
2458899.76 & r & $17.98 \pm 0.02$ \\
2458901.89 & o & $18.07 \pm 0.09$ \\
2458903.68 & r & $18.14 \pm 0.02$ \\
2458903.76 & g & $18.27 \pm 0.02$ \\
2458903.89 & c & $18.26 \pm 0.11$ \\
2458906.09 & o & $18.20 \pm 0.10$ \\
2458906.72 & r & $18.16 \pm 0.02$ \\
2458906.78 & g & $18.46 \pm 0.03$ \\
2458909.81 & r & $18.31 \pm 0.03$ \\
2458912.64 & o & $18.31 \pm 0.12$ \\
2458912.66 & g & $18.56 \pm 0.04$ \\
2458917.66 & r & $18.34 \pm 0.34$ \\
2458919.03 & o & $18.34 \pm 0.12$ \\
2458931.91 & c & $18.76 \pm 0.18$ \\
2458933.81 & o & $18.51 \pm 0.14$ \\
2458937.88 & o & $18.63 \pm 0.16$ \\
2458940.69 & g & $18.97 \pm 0.06$ \\
2458940.70 & r & $18.46 \pm 0.13$ \\
2458944.69 & r & $18.36 \pm 0.05$ \\
2458944.69 & g & $19.16 \pm 0.12$ \\
2458948.37 & o & $18.36 \pm 0.12$ \\
2458951.81 & c & $18.84 \pm 0.19$ \\
2458953.88 & o & $18.49 \pm 0.14$ \\
2458954.70 & r & $18.42 \pm 0.03$ \\
2458957.80 & o & $18.50 \pm 0.14$ \\
2458959.78 & c & $18.96 \pm 0.22$ \\
2458961.84 & o & $18.51 \pm 0.14$ \\
2458962.68 & r & $18.35 \pm 0.03$ \\
2458962.71 & g & $19.13 \pm 0.04$ \\
2458966.44 & o & $18.46 \pm 0.13$ \\
2458967.64 & r & $18.33 \pm 0.03$ \\
2458971.66 & r & $18.24 \pm 0.03$ \\
2458971.74 & g & $19.25 \pm 0.11$ \\
2458972.66 & o & $18.41 \pm 0.13$ \\
2458974.72 & r & $18.24 \pm 0.04$ \\
2458975.78 & o & $18.32 \pm 0.12$ \\
2458977.65 & g & $19.05 \pm 0.09$ \\
2458981.67 & g & $19.25 \pm 0.05$ \\
\end{tabular}
\hspace{1em}
\begin{tabular}{lcc} 
\hline\hline 
Observation &  filter & mag \\  
Date (JD) &   &  \\  \hline  
2458981.72 & r & $18.42 \pm 0.03$ \\
2458983.77 & c & $19.10 \pm 0.25$ \\
2458985.78 & o & $18.49 \pm 0.14$ \\
2458987.67 & r & $18.40 \pm 0.03$ \\
2458987.79 & c & $18.94 \pm 0.21$ \\
2458989.76 & o & $18.57 \pm 0.15$ \\
2458993.79 & o & $18.56 \pm 0.15$ \\
2458999.66 & r & $18.63 \pm 0.09$ \\
2459002.78 & o & $18.60 \pm 0.15$ \\
2459011.67 & r & $18.65 \pm 0.07$ \\
2459014.67 & r & $18.84 \pm 0.09$ \\
2459018.68 & r & $18.73 \pm 0.06$ \\
2459108.01 & r & $19.12 \pm 0.08$ \\
2459123.03 & r & $19.41 \pm 0.10$ \\
2459125.01 & g & $20.24 \pm 0.30$ \\
2459126.10 & o & $19.47 \pm 0.36$ \\
2459127.01 & r & $19.44 \pm 0.09$ \\
2459128.95 & g & $19.96 \pm 0.27$ \\
2459131.03 & r & $19.41 \pm 0.11$ \\
2459135.97 & g & $20.71 \pm 0.20$ \\
2459136.01 & r & $19.27 \pm 0.06$ \\
2459139.95 & g & $20.58 \pm 0.22$ \\
2459140.01 & r & $19.32 \pm 0.06$ \\
2459142.11 & o & $19.69 \pm 0.44$ \\
2459144.00 & r & $19.41 \pm 0.05$ \\
2459144.01 & g & $20.65 \pm 0.13$ \\
2459149.89 & r & $19.52 \pm 0.18$ \\
2459150.01 & g & $20.92 \pm 0.22$ \\
2459153.98 & r & $19.39 \pm 0.09$ \\
2459154.02 & g & $20.48 \pm 0.31$ \\
2459154.10 & o & $19.67 \pm 0.44$ \\
2459157.01 & r & $19.35 \pm 0.09$ \\
2459158.04 & o & $19.77 \pm 0.48$ \\
2459158.99 & g & $20.76 \pm 0.37$ \\
2459165.95 & g & $20.91 \pm 0.16$ \\
2459165.96 & r & $19.44 \pm 0.05$ \\
2459168.99 & g & $21.07 \pm 0.18$ \\
2459171.02 & r & $19.51 \pm 0.06$ \\
2459173.95 & g & $20.87 \pm 0.16$ \\
2459176.02 & r & $19.67 \pm 0.33$ \\
2459176.09 & c & $20.26 \pm 0.77$ \\
2459179.97 & g & $21.07 \pm 0.22$ \\
2459180.08 & o & $19.77 \pm 0.48$ \\
2459181.98 & r & $19.71 \pm 0.11$ \\
2459183.14 & o & $19.63 \pm 0.42$ \\
2459185.89 & r & $19.86 \pm 0.17$ \\
2459185.95 & g & $20.57 \pm 0.32$ \\
2459190.90 & i & $20.25 \pm 0.26$ \\
2459192.03 & o & $19.80 \pm 0.49$ \\
2459193.90 & r & $19.73 \pm 0.08$ \\
2459194.94 & g & $21.05 \pm 0.20$ \\
2459196.99 & r & $19.70 \pm 0.10$ \\
2459198.91 & g & $20.85 \pm 0.20$ \\
2459202.91 & r & $19.76 \pm 0.09$ \\
2459202.95 & i & $20.63 \pm 0.35$ \\
2459204.96 & g & $21.10 \pm 0.18$ \\
2459206.92 & r & $19.62 \pm 0.16$ \\
\end{tabular}
\hspace{1em}
\begin{tabular}{lcc} 
\hline\hline 
Observation &  filter & mag \\  
Date (JD) &   &  \\  \hline  
2459209.34 & o & $20.33 \pm 0.83$ \\
2459217.84 & r & $19.59 \pm 0.10$ \\
2459217.91 & g & $20.97 \pm 0.35$ \\
2459218.86 & o & $20.08 \pm 0.64$ \\
2459220.89 & i & $20.08 \pm 0.26$ \\
2459221.84 & g & $21.05 \pm 0.18$ \\
2459221.86 & r & $19.87 \pm 0.09$ \\
2459222.08 & o & $20.23 \pm 0.75$ \\
2459223.92 & i & $20.66 \pm 0.32$ \\
2459223.96 & c & $20.21 \pm 0.73$ \\
2459225.85 & g & $20.97 \pm 0.19$ \\
2459225.86 & r & $19.81 \pm 0.08$ \\
2459230.68 & c & $20.88 \pm 1.41$ \\
2459231.83 & r & $19.85 \pm 0.12$ \\
2459233.84 & g & $21.29 \pm 0.28$ \\
2459233.95 & c & $20.45 \pm 0.93$ \\
2459245.99 & o & $20.06 \pm 0.63$ \\
2459248.72 & g & $21.41 \pm 0.29$ \\
2459248.80 & i & $20.92 \pm 0.34$ \\
2459248.87 & r & $19.93 \pm 0.09$ \\
2459251.77 & g & $21.36 \pm 0.23$ \\
2459251.95 & r & $19.83 \pm 0.11$ \\
2459254.50 & c & $21.12 \pm 1.76$ \\
2459254.81 & g & $21.43 \pm 0.23$ \\
2459257.77 & r & $20.02 \pm 0.07$ \\
2459257.78 & i & $20.90 \pm 0.34$ \\
2459257.86 & g & $20.81 \pm 0.39$ \\
2459262.72 & r & $20.00 \pm 0.11$ \\
2459262.80 & g & $21.29 \pm 0.25$ \\
2459266.69 & i & $20.32 \pm 0.35$ \\
2459266.76 & r & $19.98 \pm 0.15$ \\
2459267.37 & o & $19.99 \pm 0.59$ \\
2459273.74 & r & $20.26 \pm 0.34$ \\
2459275.67 & g & $21.64 \pm 0.40$ \\
2459277.87 & c & $20.98 \pm 1.54$ \\
2459278.68 & r & $20.03 \pm 0.11$ \\
2459290.75 & r & $20.10 \pm 0.09$ \\
2459290.78 & i & $21.24 \pm 0.40$ \\
2459292.78 & g & $20.96 \pm 0.19$ \\
2459293.41 & o & $20.17 \pm 0.71$ \\
2459294.72 & r & $20.08 \pm 0.16$ \\
2459301.79 & r & $20.19 \pm 0.33$ \\
2459305.92 & o & $20.10 \pm 0.66$ \\
2459306.74 & r & $20.11 \pm 0.10$ \\
2459306.79 & g & $21.60 \pm 0.37$ \\
2459307.68 & i & $20.95 \pm 0.39$ \\
2459309.81 & c & $20.43 \pm 0.91$ \\
2459310.70 & r & $20.20 \pm 0.12$ \\
2459312.73 & g & $21.55 \pm 0.30$ \\
2459313.67 & i & $20.76 \pm 0.32$ \\
2459314.68 & r & $20.13 \pm 0.10$ \\
2459316.72 & g & $21.56 \pm 0.29$ \\
2459319.70 & r & $20.46 \pm 0.17$ \\
2459320.66 & o & $20.46 \pm 0.94$ \\
2459321.68 & g & $21.57 \pm 0.33$ \\
2459323.70 & r & $20.17 \pm 0.20$ \\
\end{tabular}
\end{table*}

\begin{table*}
\caption{Summary of ZTF $gri$-band and ATLAS $co$-band observations of SN 2020jfv.
ATLAS and ZTF data are binned in 3 and 0.5 days (respectively). 
The full table will be made available in machine-readable format.
} 
\label{tab:2020jfv_phot} 
\centering 
\begin{tabular}{lcc} 
\hline\hline 
Observation &  filter & mag \\  
Date (JD) &   &  \\  \hline          
2458975.11 & o & $17.59 \pm 0.06$ \\
2458999.09 & c & $18.53 \pm 0.14$ \\
2459003.10 & c & $18.71 \pm 0.17$ \\
2459005.06 & o & $18.06 \pm 0.09$ \\
2459018.89 & g & $19.45 \pm 0.07$ \\
2459018.96 & r & $18.33 \pm 0.02$ \\
2459021.05 & o & $18.44 \pm 0.13$ \\
2459023.06 & c & $18.82 \pm 0.19$ \\
2459024.88 & g & $19.21 \pm 0.08$ \\
2459024.98 & r & $18.47 \pm 0.04$ \\
2459025.05 & o & $18.51 \pm 0.14$ \\
2459027.87 & g & $19.48 \pm 0.11$ \\
2459027.98 & r & $18.60 \pm 0.05$ \\
2459029.04 & o & $18.49 \pm 0.14$ \\
2459030.86 & g & $19.48 \pm 0.18$ \\
2459030.97 & r & $18.63 \pm 0.05$ \\
2459031.04 & c & $19.01 \pm 0.23$ \\
2459033.05 & o & $18.47 \pm 0.14$ \\
2459033.86 & r & $18.61 \pm 0.07$ \\
2459033.95 & g & $19.48 \pm 0.12$ \\
2459036.88 & r & $18.57 \pm 0.07$ \\
2459036.98 & g & $19.65 \pm 0.21$ \\
2459036.99 & o & $18.54 \pm 0.14$ \\
2459038.05 & o & $18.49 \pm 0.14$ \\
2459042.96 & r & $18.61 \pm 0.04$ \\
2459043.09 & o & $18.55 \pm 0.15$ \\
2459045.04 & o & $18.84 \pm 0.19$ \\
2459045.86 & r & $18.73 \pm 0.04$ \\
2459045.95 & g & $19.59 \pm 0.07$ \\
2459048.90 & r & $18.78 \pm 0.03$ \\
2459049.04 & o & $18.86 \pm 0.20$ \\
2459051.03 & c & $19.39 \pm 0.33$ \\
2459051.92 & r & $18.82 \pm 0.03$ \\
2459051.96 & g & $19.76 \pm 0.07$ \\
2459053.03 & o & $19.03 \pm 0.23$ \\
2459054.96 & g & $19.67 \pm 0.06$ \\
2459057.95 & r & $18.85 \pm 0.04$ \\
2459059.00 & c & $19.25 \pm 0.29$ \\
2459060.92 & r & $18.95 \pm 0.03$ \\
2459060.96 & o & $18.97 \pm 0.22$ \\
2459061.85 & r & $18.87 \pm 0.05$ \\
2459062.47 & o & $18.90 \pm 0.21$ \\
2459062.89 & g & $19.84 \pm 0.17$ \\
2459064.93 & r & $19.00 \pm 0.10$ \\
2459064.97 & o & $19.12 \pm 0.25$ \\
2459064.98 & g & $19.66 \pm 0.24$ \\
2459069.88 & g & $19.92 \pm 0.15$ \\
2459069.90 & r & $18.93 \pm 0.06$ \\
2459071.72 & o & $19.06 \pm 0.24$ \\
2459072.86 & g & $19.83 \pm 0.11$ \\
2459075.02 & o & $19.19 \pm 0.27$ \\
2459075.81 & r & $19.12 \pm 0.05$ \\
2459075.86 & g & $19.71 \pm 0.07$ \\
2459077.06 & o & $19.07 \pm 0.24$ \\
2459077.91 & g & $19.80 \pm 0.08$ \\
2459078.85 & r & $19.10 \pm 0.04$ \\
2459078.86 & g & $19.82 \pm 0.08$ \\
2459078.97 & c & $19.31 \pm 0.31$ \\
2459079.91 & r & $19.02 \pm 0.04$ \\
2459080.98 & o & $19.14 \pm 0.26$ \\
2459081.84 & g & $19.92 \pm 0.10$ \\
2459081.88 & r & $19.12 \pm 0.05$ \\
2459082.89 & g & $19.13 \pm 0.15$ \\
2459082.94 & c & $19.73 \pm 0.46$ \\
\end{tabular}
\hspace{1em}
\begin{tabular}{lcc} 
\hline\hline 
Observation &  filter & mag \\  
Date (JD) &   &  \\  \hline  
2459084.88 & g & $19.96 \pm 0.09$ \\
2459084.94 & o & $19.14 \pm 0.26$ \\
2459085.92 & g & $19.83 \pm 0.08$ \\
2459086.79 & g & $19.88 \pm 0.08$ \\
2459086.87 & r & $19.14 \pm 0.05$ \\
2459087.04 & c & $19.83 \pm 0.51$ \\
2459089.83 & g & $20.23 \pm 0.20$ \\
2459089.99 & o & $19.21 \pm 0.28$ \\
2459090.85 & g & $20.06 \pm 0.22$ \\
2459091.82 & g & $20.20 \pm 0.25$ \\
2459092.90 & g & $19.55 \pm 0.18$ \\
2459098.83 & r & $19.12 \pm 0.08$ \\
2459099.01 & o & $19.29 \pm 0.30$ \\
2459099.85 & r & $19.22 \pm 0.08$ \\
2459102.10 & o & $19.17 \pm 0.27$ \\
2459104.94 & o & $19.29 \pm 0.30$ \\
2459105.89 & c & $19.78 \pm 0.48$ \\
2459106.93 & c & $19.65 \pm 0.43$ \\
2459107.80 & r & $19.17 \pm 0.07$ \\
2459107.82 & g & $20.02 \pm 0.13$ \\
2459108.90 & o & $19.39 \pm 0.33$ \\
2459110.78 & r & $19.21 \pm 0.10$ \\
2459110.82 & g & $19.92 \pm 0.12$ \\
2459110.88 & c & $20.10 \pm 0.66$ \\
2459112.87 & o & $19.34 \pm 0.31$ \\
2459113.76 & r & $19.15 \pm 0.05$ \\
2459114.89 & c & $19.80 \pm 0.49$ \\
2459116.80 & g & $19.94 \pm 0.08$ \\
2459116.82 & r & $19.23 \pm 0.06$ \\
2459117.80 & g & $19.81 \pm 0.08$ \\
2459117.87 & o & $19.33 \pm 0.31$ \\
2459118.77 & r & $19.15 \pm 0.07$ \\
2459118.86 & o & $19.11 \pm 0.25$ \\
2459120.84 & g & $20.31 \pm 0.35$ \\
2459124.75 & r & $19.40 \pm 0.17$ \\
2459126.92 & o & $19.35 \pm 0.32$ \\
2459128.77 & r & $19.29 \pm 0.08$ \\
2459128.78 & g & $19.94 \pm 0.17$ \\
2459129.38 & o & $19.40 \pm 0.33$ \\
2459130.74 & g & $19.98 \pm 0.10$ \\
2459130.76 & r & $19.20 \pm 0.06$ \\
2459131.75 & o & $19.50 \pm 0.37$ \\
2459134.71 & r & $19.38 \pm 0.06$ \\
2459134.73 & g & $20.21 \pm 0.12$ \\
2459135.02 & o & $19.25 \pm 0.29$ \\
2459136.69 & r & $19.21 \pm 0.05$ \\
2459136.75 & g & $20.10 \pm 0.10$ \\
2459136.89 & o & $19.47 \pm 0.36$ \\
2459138.78 & r & $19.33 \pm 0.08$ \\
2459138.88 & c & $20.04 \pm 0.62$ \\
2459140.69 & g & $20.10 \pm 0.10$ \\
2459140.73 & r & $19.28 \pm 0.05$ \\
2459140.84 & o & $19.44 \pm 0.35$ \\
2459142.84 & c & $19.85 \pm 0.52$ \\
2459144.82 & o & $19.24 \pm 0.28$ \\
2459145.68 & r & $19.34 \pm 0.06$ \\
2459145.73 & g & $20.02 \pm 0.10$ \\
2459146.87 & o & $19.47 \pm 0.36$ \\
2459147.77 & r & $19.17 \pm 0.30$ \\
2459151.71 & r & $19.39 \pm 0.13$ \\
2459153.61 & r & $19.34 \pm 0.12$ \\
2459153.69 & g & $19.74 \pm 0.21$ \\
2459155.67 & g & $20.25 \pm 0.22$ \\
2459155.75 & r & $19.15 \pm 0.08$ \\
\end{tabular}
\hspace{1em}
\begin{tabular}{lcc} 
\hline\hline 
Observation &  filter & mag \\  
Date (JD) &   &  \\  \hline  
2459155.86 & o & $19.31 \pm 0.31$ \\
2459156.70 & r & $19.43 \pm 0.13$ \\
2459158.64 & r & $19.23 \pm 0.06$ \\
2459158.67 & g & $20.26 \pm 0.18$ \\
2459158.79 & o & $19.06 \pm 0.24$ \\
2459160.85 & o & $19.22 \pm 0.28$ \\
2459164.71 & g & $19.99 \pm 0.17$ \\
2459167.67 & g & $19.99 \pm 0.12$ \\
2459167.71 & r & $19.19 \pm 0.06$ \\
2459169.65 & r & $19.19 \pm 0.08$ \\
2459169.71 & g & $20.24 \pm 0.16$ \\
2459170.79 & c & $20.42 \pm 0.90$ \\
2459171.65 & g & $20.01 \pm 0.16$ \\
2459173.61 & r & $19.16 \pm 0.06$ \\
2459173.63 & o & $19.29 \pm 0.30$ \\
2459173.75 & g & $20.03 \pm 0.13$ \\
2459178.69 & r & $18.97 \pm 0.08$ \\
2459180.62 & r & $19.00 \pm 0.13$ \\
2459180.77 & o & $19.17 \pm 0.27$ \\
2459182.61 & r & $19.08 \pm 0.09$ \\
2459182.65 & g & $19.82 \pm 0.20$ \\
2459182.78 & o & $19.21 \pm 0.28$ \\
2459184.61 & r & $19.08 \pm 0.08$ \\
2459184.66 & g & $20.43 \pm 0.34$ \\
2459185.81 & o & $19.21 \pm 0.28$ \\
2459186.65 & g & $19.95 \pm 0.24$ \\
2459186.72 & i & $19.08 \pm 0.21$ \\
2459189.61 & g & $20.09 \pm 0.12$ \\
2459189.70 & r & $19.02 \pm 0.07$ \\
2459190.75 & c & $19.93 \pm 0.56$ \\
2459192.63 & r & $19.18 \pm 0.35$ \\
2459192.70 & g & $19.68 \pm 0.30$ \\
2459192.78 & o & $19.25 \pm 0.29$ \\
2459194.64 & r & $19.08 \pm 0.06$ \\
2459194.69 & g & $20.20 \pm 0.19$ \\
2459194.76 & c & $19.86 \pm 0.52$ \\
2459195.65 & g & $20.21 \pm 0.26$ \\
2459198.73 & c & $20.03 \pm 0.62$ \\
2459199.61 & r & $19.13 \pm 0.07$ \\
2459199.69 & g & $20.21 \pm 0.23$ \\
2459202.65 & g & $20.04 \pm 0.22$ \\
2459206.71 & o & $19.53 \pm 0.38$ \\
2459208.73 & o & $19.27 \pm 0.29$ \\
2459210.60 & r & $19.17 \pm 0.12$ \\
2459211.67 & i & $19.90 \pm 0.34$ \\
2459212.71 & o & $19.37 \pm 0.32$ \\
2459215.58 & r & $19.29 \pm 0.12$ \\
2459216.07 & o & $19.15 \pm 0.26$ \\
2459217.61 & r & $19.13 \pm 0.26$ \\
2459218.75 & c & $19.71 \pm 0.45$ \\
2459219.61 & g & $20.24 \pm 0.15$ \\
2459219.63 & r & $19.05 \pm 0.06$ \\
2459219.66 & i & $20.09 \pm 0.35$ \\
2459222.24 & c & $19.96 \pm 0.58$ \\
2459223.63 & r & $19.03 \pm 0.07$ \\
2459223.63 & g & $20.17 \pm 0.19$ \\
2459224.64 & i & $19.93 \pm 0.34$ \\
2459224.74 & c & $20.28 \pm 0.79$ \\
2459227.63 & r & $19.02 \pm 0.10$ \\
2459230.60 & r & $19.08 \pm 0.09$ \\
2459231.60 & r & $18.98 \pm 0.12$ \\
2459232.60 & r & $18.92 \pm 0.07$ \\
2459233.60 & r & $19.13 \pm 0.09$ \\
2459234.72 & o & $19.30 \pm 0.30$ \\
\end{tabular}
\end{table*}

\begin{deluxetable}{lcc}
\tablewidth{0pt}
\tabletypesize{\scriptsize}
\tablecaption{Summary of PGIR $J$-band observations of  SN 2020jfo.$^a$ 
\label{tab:pgir}}
\tablehead{
\colhead{Observation Date} & 
\colhead{Rest-Frame Phase} & 
\colhead{$J$} \\
\colhead{(JD)} &
\colhead{(days)} &
\colhead{(mag)}
}
\startdata
2458635.71 & -328.22 & $>14.87$ \\
2458640.73 & -323.28 & $>15.17$ \\
2458648.71 & -315.44 & $>15.12$ \\
2458669.68 & -294.82 & $>14.84$ \\
2458676.68 & -287.94 & $>14.87$ \\
2458682.67 & -282.05 & $>14.84$ \\
2458866.99 & -100.81 & $>14.82$ \\
2458896.06 & -72.22 & $>14.98$ \\
2458913.02 & -55.55 & $>14.82$ \\
2458964.87 & -4.56 & $>14.98$ \\
2458967.87 & -1.61 & $>14.88$ \\
2458976.81 & 7.18 & $15.10 \pm 0.23$ \\
2458977.83 & 8.18 & $14.64 \pm 0.18$ \\
2458980.74 & 11.04 & $14.29 \pm 0.10$ \\
2458986.74 & 16.94 & $14.15 \pm 0.10$ \\
2458993.75 & 23.83 & $14.02 \pm 0.10$ \\
2458995.68 & 25.73 & $13.96 \pm 0.08$ \\
2458997.23 & 27.26 & $14.16 \pm 0.09$ \\
2459004.77 & 34.67 & $14.15 \pm 0.19$ \\
2459011.75 & 41.53 & $14.19 \pm 0.27$ \\
2459019.72 & 49.37 & $14.10 \pm 0.23$ \\
2459177.06 & 204.08 & $>14.79$ \\
2459223.00 & 249.25 & $>14.18$ \\
2459249.93 & 275.73 & $>14.49$ \\
2459269.87 & 295.34 & $>14.86$ \\
2459280.84 & 306.13 & $>15.17$ \\
\enddata
\tablenotetext{a}{Fluxes with S/N $< 3\sigma$ are shown as upper limits. These magnitudes are Vega tied to 2MASS, and $m$(AB) $-$ $m$(Vega) = 0.91.
}
\end{deluxetable}

\clearpage

\begin{deluxetable}{lcccccccc}
\tablewidth{0pt}
\tabletypesize{\scriptsize}
\tablecaption{Summary of {\it Swift}/UVOT observations of  SN 2020jfo.$^a$ 
%Fluxes with SNR less than 3 sigma are shown as upper limits.
\label{tab:swift}}
\tablehead{
\colhead{Observation Date} & 
\colhead{Rest-Frame Phase} & 
\colhead{$V$} & 
\colhead{$B$} & 
\colhead{$U$} &
\colhead{$UVW1$} &
\colhead{$UVW2$} &
\colhead{$UVM2$} \\
\colhead{(MJD)} &
\colhead{(days)}  & 
\colhead{(mag)} &
\colhead{(mag)} &
\colhead{(mag)} &
\colhead{(mag)} &
\colhead{(mag)} &
\colhead{(mag)} 
}
\startdata
2458976.62 & 1.41 & 15.00 (0.06) & 14.77 (0.04) & 14.57 (0.05) & 14.69 (0.05) & 14.79 (0.06) & 14.78 (0.05) \\
2458978.23 & 3.02 & 14.66 (0.06) & 14.57 (0.05) & 14.41 (0.05) & 14.65 (0.05) & 14.85 (0.06) & 14.71 (0.05) \\
2458978.73 & 3.51 & 14.66 (0.06) & 14.61 (0.05) & 14.43 (0.05) & 14.83 (0.05) & 15.21 (0.07) & 14.89 (0.05) \\
2458979.85 & 4.63 & 14.55 (0.06) & 14.51 (0.05) & 14.41 (0.05) & 14.90 (0.05) & 15.41 (0.07) & 15.12 (0.05) \\
2458979.92 & 4.70 & 14.63 (0.07) & 14.59 (0.05) & 14.51 (0.05) & 15.03 (0.05) & 15.51 (0.07) & 15.17 (0.05) \\
2458980.72 & 5.49 & 14.54 (0.05) & 14.51 (0.05) & 14.41 (0.05) & 14.98 (0.05) & 15.79 (0.07) & 15.38 (0.05) \\
2458982.26 & 7.02 & 14.62 (0.06) & 14.63 (0.05) & 14.57 (0.05) & 15.42 (0.05) & 16.37 (0.07) & 15.96 (0.05) \\
2458982.92 & 7.68 & 14.61 (0.06) & 14.55 (0.05) & 14.58 (0.05) & 15.51 (0.05) & 16.38 (0.07) & 16.09 (0.06) \\
2458985.17 & 9.92 & 14.77 (0.06) & 14.61 (0.05) & 14.81 (0.05) & 15.99 (0.06) & 16.88 (0.08) & 16.78 (0.06) \\
2458986.16 & 10.91 & 14.78 (0.08) & 14.68 (0.06) & 14.85 (0.06) & 16.21 (0.07) & 17.18 (0.09) & 17.08 (0.08) \\
2458994.41 & 19.11 & 14.69 (0.06) & 15.00 (0.05) & 16.27 (0.07) & 18.24 (0.12) & 19.96 (0.27) & 20.03 (0.27) \\
2458996.06 & 20.75 & 14.72 (0.06) & 15.09 (0.06) & 16.63 (0.08) & 18.56 (0.14) & 19.79 (0.24) & 20.41 (0.36) \\
2459003.26 & 27.91 & 14.82 (0.05) & 15.49 (0.05) & 17.33 (0.08) & 19.27 (0.17) & $>$ 20.66 & $>$ 20.80 \\
2459007.61 & 32.24 & 14.89 (0.05) & 15.61 (0.05) & 17.70 (0.09) & 19.54 (0.19) & 20.61 (0.33) & $>$ 20.85 \\
2459013.59 & 38.19 & 14.97 (0.05) & 15.76 (0.05) & 18.08 (0.11) & 19.92 (0.26) & $>$ 20.74 & $>$ 20.84 \\
2459018.49 & 43.07 & 14.91 (0.07) & 15.89 (0.08) & 18.55 (0.24) & 19.61 (0.32) & $>$ 20.43 & $>$ 20.36 \\
2459048.91 & 73.33 & 17.45 (0.14) & 18.44 (0.19) & $>$ 20.05 & $>$ 20.62 & $>$ 20.87 & $>$ 21.00 \\
2459059.83 & 84.19 & 17.29 (0.13) & 18.64 (0.23) & $>$ 20.02 & $>$ 20.58 & $>$ 20.88 & $>$ 20.99 \\
2459064.91 & 89.24 & 17.55 (0.16) & 18.86 (0.27) & $>$ 20.03 & $>$ 20.59 & $>$ 20.90 & $>$ 20.95 \\
2459070.00 & 94.31 & 17.46 (0.20) & 18.75 (0.25) & $>$ 20.01 & $>$ 20.61 & $>$ 20.89 & $>$ 20.81 \\
2459168.08 & 191.88 & $>$ 18.06 & $>$ 19.08 & $>$ 19.85 & $>$ 20.49 & $>$ 20.72 & $>$ 20.77 \\
2459195.53 & 219.19 & $>$ 18.52 & $>$ 19.19 & $>$ 19.91 & $>$ 20.54 & $>$ 20.82 & $>$ 20.91 \\
2459207.91 & 231.50 & $>$ 18.81 & $>$ 19.43 & $>$ 20.15 & $>$ 20.67 & $>$ 20.95 & $>$ 21.02 \\
2459217.75 & 241.29 & 18.73 (0.34) & $>$ 19.46 & $>$ 20.18 & $>$ 20.64 & $>$ 20.99 & $>$ 21.04 \\
2459228.02 & 251.51 & $>$ 18.84 & $>$ 19.45 & $>$ 20.21 & $>$ 20.66 & $>$ 20.99 & $>$ 21.07 \\
\enddata
\tablenotetext{a}{Fluxes with S/N $< 3\sigma$ are shown as upper limits.}
\end{deluxetable}

\clearpage

\begin{deluxetable}{lccc}
%\begin{table*}{lccc}
\tablewidth{0pt}
\tabletypesize{\scriptsize}
\tablecaption{Summary of Spectroscopic Observations \label{tab:spec}}
\tablehead{
\colhead{Object} &
\colhead{UT Observation Date} & 
\colhead{Rest-Frame Phase} & % this is rest frame
\colhead{Telescope+Instrument} \\
\colhead{} &
\colhead{(YYYY MM DD)}  & 
\colhead{(days)} &
\colhead{} 
}
\startdata
SN\,2020jfo & 2020 May 06 & 1.21 & LT+SPRAT \\
SN\,2020jfo & 2020 May 06 & 1.22 & NOT+ALFOSC \\
SN\,2020jfo & 2020 May 07 & 1.66 & P60+SEDM \\
SN\,2020jfo & 2020 May 07 & 2.22 & LT+SPRAT\\
SN\,2020jfo & 2020 May 08 & 3.22 & LT+SPRAT\\  
SN\,2020jfo & 2020 May 09 & 3.49 & P60+SEDM \\
SN\,2020jfo & 2020 May 10 & 4.64 & P60+SEDM \\
SN\,2020jfo & 2020 May 15 & 9.56 & P60+SEDM \\
SN\,2020jfo & 2020 May 17 & 11.50 & Lick+Kast \\
SN\,2020jfo & 2020 May 25 & 19.48 & Lick+Kast \\
SN\,2020jfo & 2020 May 25 & 19.53 & P60+SEDM \\
SN\,2020jfo & 2020 May 29 & 23.44 & Lick+Kast \\
SN\,2020jfo & 2020 Jun 12 & 37.31 & P60+SEDM \\
SN\,2020jfo & 2020 Jun 17 & 42.36 & P60+SEDM \\
SN\,2020jfo & 2020 Dec 04 & 211.63 & P60+SEDM \\
SN\,2020jfo & 2020 Dec 07 & 214.40 & NOT+ALFOSC \\
SN\,2020jfo & 2020 Dec 22 & 229.63 & P60+SEDM \\
SN\,2020jfo & 2021 Jan 16 & 254.17 & NOT+ALFOSC \\
SN\,2020jfo & 2021 Feb 04 & 273.33 & P60+SEDM \\
SN\,2020jfo & 2021 Feb 09 & 277.93 & NOT+ALFOSC \\
SN\,2020jfo & 2021 Mar 09 & 305.77 & NOT+ALFOSC \\
SN\,2020jfo & 2021 Apr 20 & 348.41 & NOT+ALFOSC \\
\hline
SN\,2020amv & 2020 Jan 24 & 1.67 & P60+SEDM \\
SN\,2020amv & 2020 Jan 28 & 5.42 & P60+SEDM \\
SN\,2020amv & 2020 Jan 29 & 6.59 & GeminiN+GMOS \\
SN\,2020amv & 2020 Jan 31 & 8.12 & P60+SEDM \\
SN\,2020amv & 2020 Feb 02 & 9.84 & NTT+EFOSC2\tablenotemark{a}\\
SN\,2020amv & 2020 Feb 04 & 12.03 & P60+SEDM \\
SN\,2020amv & 2020 Feb 06 & 13.89 & P60+SEDM \\
SN\,2020amv & 2020 Sep 19 & 230.37 & P60+SEDM \\
SN\,2020amv & 2020 Sep 27 & 238.00 & P60+SEDM \\
SN\,2020amv & 2020 Nov 18 & 287.67 & P200+DBSP \\
SN\,2020amv & 2021 Jan 23 & 350.61 & NOT+ALFOSC \\
SN\,2020amv & 2021 Feb 14 & 372.35 & NOT+ALFOSC \\
SN\,2020amv & 2021 Apr 07 & 421.58 & Keck+LRIS \\
\hline
SN\,2020jfv & 2020 Jun 20 & 50.55 & P60+SEDM \\
SN\,2020jfv & 2020 Jun 24 & 54.42 & P60+SEDM \\
SN\,2020jfv & 2020 Dec 01 & 211.58 & P60+SEDM \\
SN\,2020jfv & 2020 Dec 06 & 217.18 & NOT+ALFOSC \\
SN\,2020jfv & 2020 Dec 07 & 217.45 & P60+SEDM \\
SN\,2020jfv & 2021 Jan 14 & 254.89 & Keck+LRIS \\
{  SN\,2020jfv} & {  2021 Jul 08} & {  426.89} & {  NOT+ALFOSC} \\
\enddata
\tablenotetext{a}{Uploaded from TNS from ePESSTO.}
\end{deluxetable}
%\end{table*}

\begin{table*}
\caption{Supernova light-curve properties}              % title of Table
\label{table:SNLCproperties}       % is used to refer this table in the text
\centering                                      % used for centering table
\begin{tabular}{lccccccccc}          % centered columns (4 columns)
\hline\hline                        % inserts double horizontal lines
SN Name &  $t^{\rm{rise}}_{g}$ & $t^{\rm{rise}}_{r}$ & $t^{\rm{rise}}_{i}$ &
$m^{\rm{peak}}_{g}$ &
$m^{\rm{peak}}_{r}$ &
$m^{\rm{peak}}_{i}$ &
$M^{\rm{peak}}_{g}$ &
$M^{\rm{peak}}_{r}$ &
$M^{\rm{peak}}_{i}$ \\
&   & (rest-frame days past explosion) &  & &(mag) & && (mag) & \\    
% table heading
\hline                                   % inserts single horizontal line
SN 2020amv & 8.99 & 14.16 & 16.46 & 17.38 & 17.57 & 17.62 & -19.23 & -19.04 & -18.99 \\    
SN 2020jfo & 4.68 & 4.87 & 5.97  & 14.38 & 14.37 & 14.65 & -16.49 & -16.50  & -16.22 \\    
SN 2020jfv & -- & -- & -- & $<18.53$ & 17.71 & 17.28 & -15.95 & -16.77 & -17.22 \\ 
\hline                                             
\end{tabular}
\end{table*}

\end{document}